\documentclass[lettersize,journal]{IEEEtran}

\usepackage{multirow}
\usepackage{makecell}

\usepackage{amsmath}
\usepackage{booktabs}
\usepackage{multirow}
\usepackage{color}
\usepackage{hyperref}
\usepackage{breakurl} 
\usepackage{graphicx}
\usepackage{epstopdf}
\usepackage{subfigure}
\usepackage{float}

\usepackage{macro}

\usepackage{cite}
\usepackage{amsmath}
\usepackage{amsfonts}
\usepackage{amsmath,amssymb,amsfonts}

\usepackage{algorithm}  
\usepackage{algpseudocode}

\usepackage{caption}

\usepackage{graphicx}
\usepackage{textcomp}
\usepackage{xcolor}
\usepackage{flushend}

\begin{document}

\title{Code Copycat Conundrum: Demystifying Repetition in LLM-based Code Generation}

\author{
Mingwei~Liu, 
Juntao~Li, 
Ying~Wang, 
Xueying~Du, 
Zuoyu~Ou, 
Qiuyuan~Chen, 
Bingxu~An, 
Zhao~Wei, 
Yong~Xu, 
Fangming~Zou, 
Xin~Peng, 
Yiling~Lou
\thanks{Mingwei Liu is with the Sun Yat-sen University, Zhuhai, China (e-mail:
liumw26@mail.sysu.edu.cn).}
\thanks{Juntao Li, Ying Wang, Xueying Du, Zuoyu Ou, Xin Peng, and Yiling Lou
are with the Fudan University, Shanghai, China (e-mail: 22210240197@
fudan.edu.cn; 22210240051@fudan.edu.cn; 21210240012@m.fudan.edu.cn; 22210240250@fudan.edu.cn;  pengxin@fudan.edu.cn; yilinglou@fudan.edu.cn).}%
\thanks{Qiuyuan Chen, Bingxu An, Zhao Wei, Yong Xu, and Fangming Zou
are with the Tencent Technology (e-mail: joeqychen@tencent.com; bingxuan@tencent.com; zachwei@tencent.com; rogerxu@tencent.com; stephenzou@tencent.com).}}

\markboth{Journal of \LaTeX\ Class Files,~Vol.~14, No.~8, August~2021}%
{Code Copycat Conundrum: Demystifying Repetition in LLM-based Code Generation}

\IEEEpubid{0000--0000/00\$00.00~\copyright~2021 IEEE}

\maketitle

\begin{abstract}
Despite recent advances in Large Language Models (LLMs) for code generation, the quality of LLM-generated code still faces significant challenges. One significant issue is code repetition, which refers to the model's tendency to generate structurally redundant code, resulting in inefficiencies and reduced readability. To address this, we conduct the first empirical study to investigate the prevalence and nature of repetition across 19 state-of-the-art code LLMs using three widely-used benchmarks. Our study includes both quantitative and qualitative analyses, revealing that repetition is pervasive and manifests at various granularities and extents, including character, statement, and block levels. We further summarize a taxonomy of 20 repetition patterns. 

Building on our findings, we propose \app{}, a rule-based technique designed to detect and mitigate repetition in generated code. We evaluate \app{} using both open-source benchmarks and in an industrial setting. Our results demonstrate that \app{} significantly outperforms baselines in reducing repetition (with an average improvements of 91.3\%, 93.5\%, and 79.9\% in rep-3, rep-line, and sim-line metrics) and enhancing code quality (with a Pass@1 increase of 208.3\% over greedy search). Furthermore, integrating \app{} improves the performance of existing repetition mitigation methods, with Pass@1 improvements ranging from 53.7\% to 215.7\%.

\end{abstract}

\begin{IEEEkeywords}
Code Search, Decoder-only LLMs
\end{IEEEkeywords}

\section{Introduction}

\IEEEPARstart{T}{he} recent advance of Large Language Models (LLMs) has significantly boosted code generation techniques. State-of-the-art code LLMs (e.g., StarCoder~\cite{li2023starcoder}, CodeLlama~\cite{DBLP:journals/corr/abs-2308-12950}, and DeepSeek-Coder~\cite{deepseek-coder}) demonstrate remarkable effectiveness in generating code for the given natural language descriptions, exhibit remarkable effectiveness in generating code from natural language descriptions, owing to their pre-training on extensive textual and code corpora. However, despite these advancements, the quality of code generated by LLMs still faces several challenges, which hinder its widespread application~\cite{DBLP:conf/icse/FanGMRT23}. 

\begin{figure*}[htb]
	\centering
        \vspace{-1mm}
	\includegraphics[width=1.4\columnwidth]{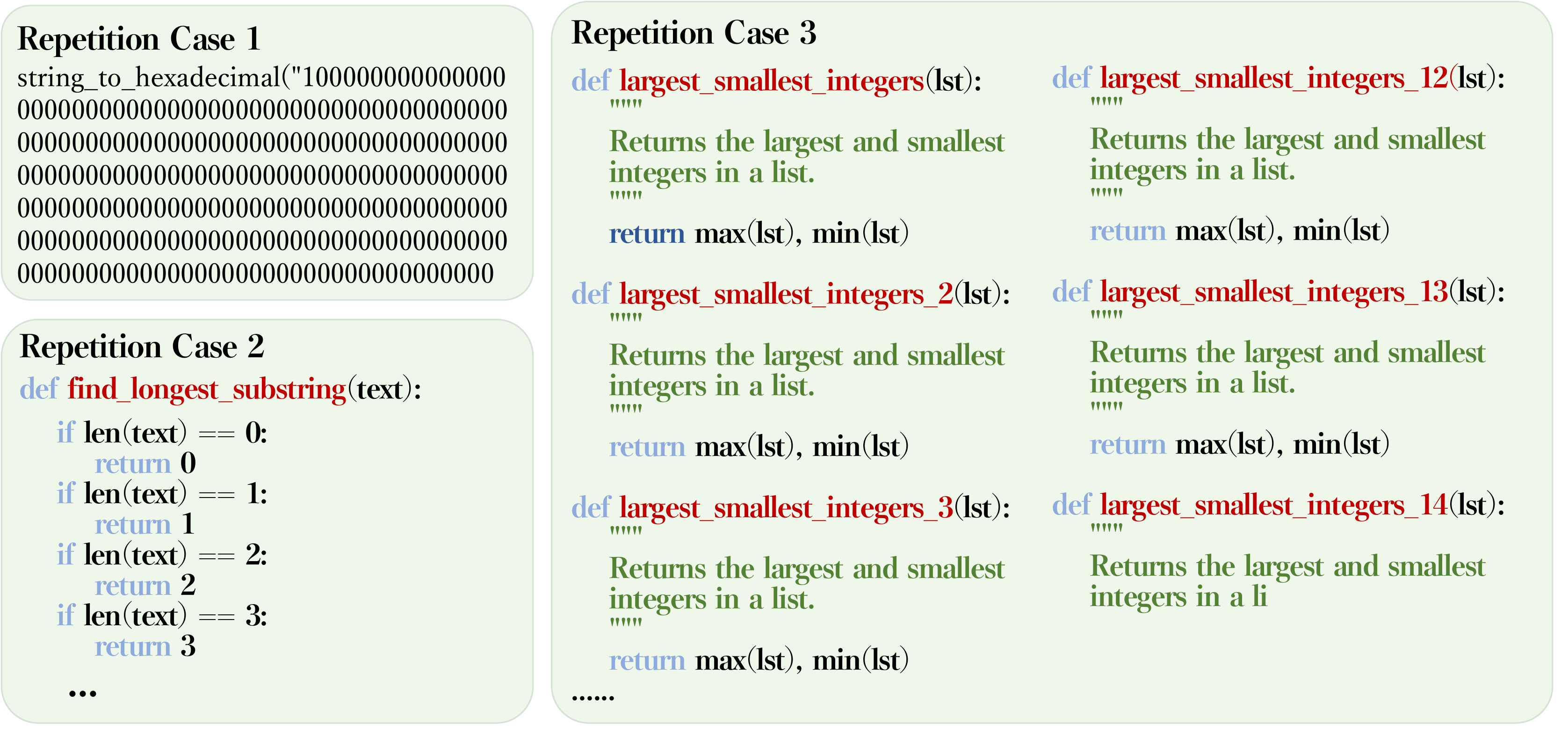}
	\caption{Examples of Endless Repetition Cases of Code LLMs}
	\label{fig:endless repetition cases}
        \vspace{-1mm}
\end{figure*}

In this work, we focus on repetition, a significant challenge in LLM-based code generation. Repetition refers to the model's tendency to generate structurally or textually redundant code. This includes exact duplicates and near-duplicates, where code fragments differ only slightly, such as in variable names or constant values. It also includes meaningless character-level repetition, such as sequences like ``12345671234567...'' or long runs of repeated digits like ``000000...''. Figure~\ref{fig:endless repetition cases} presents three representative types: character repetition, where the same character is repeated continuously as shown in Case 1; statement repetition, where similar statements are generated repeatedly as seen in Case 2; and function repetition, where the same function is defined multiple times as illustrated in Case 3. These types of repetition significantly reduce code quality and readability, and can undermine the practical usability of automatic code generation tools. Our preliminary analysis revealed 10,399 code snippets containing repetition, among which 9,346 cases (89.9\%) exceeded the predefined maximum token limit, resulting in truncated and incomplete outputs.

Existing researches have explored similar repetition issues in text generation, commonly referred to as neural text degeneration. Li et al.~\cite{DBLP:conf/nips/LiLF0LCWS23} highlight the correlation between repetitive outputs and repetitive training data, and Xu et al.~\cite{DBLP:conf/nips/XuLY0LL22} propose solutions to mitigate the issue. However, to date, there is no investigation into repetition issues specifically within code generation tasks, where the inherently structured and repetitive nature of programming further exacerbates the issue.

\textbf{Empirical Study.} To fill this gap, we perform the first study of repetition issues in LLM-based code generation. In particular,  (i) we  quantitatively analyze the prevalence of repetition issues across 19 state-of-the-art code LLMs on three widely-used code generation benchmarks, and  (ii) we qualitatively summarize the taxonomy of 20 repetition patterns. Overall, we find that repetition is indeed pervasive in LLM-based code generation; and our distilled repetition patterns involve different repetitive granularities (i.e., character level, statement level, and block level) and different repetitive extent (i.e., complete, similar, finite, infinite, or random repetition). 

\textbf{Repetition Mitigation Technique.} Inspired by our empirical studies above, we further propose a rule-based repetition mitigation technique, \app{}, which first detects repetition issues of different granularities in LLM-generated code and then fixes the identified repetition respectively. 

\textbf{Open-Source and Industrial Evaluation.} We further  evaluate the effectiveness and efficiency of \app{} in both the open-source code generation benchmark and the industrial code generation within the company A (anonymous for the double-blind policy). The results of both settings consistently demonstrate the substantial improvements of \app{} in precisely reducing repetition issues and enhancing code quality. 

In summary, this work makes the following contributions:
\begin{itemize}[itemsep=2pt,topsep=0pt,parsep=0pt]

    \item \textbf{The first empirical study} of repetition in LLM-based code generation, including both quantitative analysis of the prevalence of repetition issues and qualitative analysis of the 20 recurring repetition patterns.
    
    \item \textbf{The first repetition mitigation technique \app{}}, which automatically detects and fixes the repetition in LLM-generated code. 
    
\IEEEpubidadjcol
    \item \textbf{A comprehensive evaluation} on both open-source code generation benchmark and real-world industrial code generation, showing the effectiveness of \app{} in precisely reducing the repetition issues. All code and data is included in our online  replication package~\cite{decoder2024}.

\end{itemize}

\section{Related Work}
Code generation involves producing code snippets from natural language descriptions and has been extensively studied~\cite{vikram2023large,10172763,kang2023explainable}. Recent advancements in LLMs, such as StarCoder~\cite{li2023starcoder}, CodeLlama~\cite{DBLP:journals/corr/abs-2308-12950}, and DeepSeek-Coder~\cite{deepseek-coder}, have enhanced these capabilities, leveraging large code-specific corpora.
To evaluate the performance of these code LLMs, several high-quality benchmarks have been introduced. Notable examples include HumanEval, MBPP, TestEval, BigCodeBench, CrossCodeEval, EvoCodeBench and ClassEval, which cover a range of tasks from repository-level and class-level code generation to test case generation~\cite{chen2021huamneval,repocoder,austin2021mbpp,wang2024testeval,zhuo2024bigcodebench,classeval,ding2024crosscodeeval,li2024evocodebench}.
Recent research has explored issues in code generated by LLMs such as bugs~\cite{tambon2024bugs}, hallucinations~\cite{liu2024exploring}, and coding style inconsistencies~\cite{wang2024beyond}. This paper is the first to systematically study repetition in code LLMs, analyzing 19 models and categorizing repetition patterns. 

Repetition in LLM-generated text is well-documented, with Li et al.~\cite{DBLP:conf/nips/LiLF0LCWS23} linking it to training data characteristics. In code generation, repetition results in inefficient and redundant code, impacting performance and readability. Chen et al.~\cite{chen2021huamneval} find that LLMs often produce repetitive patterns due to overfitting to common structures. Current detection methods include n-gram overlap and AST comparison, while strategies like DITTO~\cite{DBLP:conf/nips/XuLY0LL22} focus on reducing repetition through various techniques. Our empirical findings lead to the development of DeRep, a rule-based approach tailored for code generation that significantly improves upon general techniques by addressing the specific challenges of code repetition.

\section{Empirical Study Setup}
We empirically investigate repetition issues in LLM-based code generation by addressing the following two RQs.

\begin{itemize}[leftmargin=15pt]
\item \textbf{RQ1 (Quantitative analysis): how prevalent are the repetition issues in LLM-based code generation?} In this RQ, we automatically evaluate the prevalence of repetition issues in a range of state-of-the-art code LLMs on widely-used code generation benchmarks. 

\begin{itemize}[leftmargin=15pt]
\item \textbf{RQ1.a: how are the repetition issues in terms of different metrics?}
\item \textbf{RQ1.b: how are the repetition issues across different code LLMs?}
\item \textbf{RQ1.c: how are the repetition issues across different coding tasks?}
\end{itemize}

\item \textbf{RQ2 (Qualitative analysis): what are the recurring repetition patterns in LLM-based code generation?} In this RQ, we manually inspect and summarize the recurring repetition patterns in the code generated by the studied code LLMs.  
\end{itemize}

\begin{table}[]
\footnotesize
	\centering
   \caption{Studied Code LLMs}
\label{table:baseline_info}
\begin{tabular}{l|l|l|l|r}
\hline
\textbf{Model}                                                                               & \textbf{Base Model}                                                                   & \textbf{Instruct} & \textbf{Time} & \textbf{Size} \\ \hline
\multicolumn{1}{l|}{SantaCoder\cite{allal2023santacoder}}                   & \multicolumn{1}{l|}{}                                                                 & n                 & 2023.1        & 1.1B          \\ \hline
\multicolumn{1}{l|}{StarCoder\cite{li2023starcoder}}                        & \multicolumn{1}{l|}{}                                                                 & n                 & 2023.5        & 15.5B         \\ \hline
\multicolumn{1}{l|}{\multirow{3}{*}{StarCoder2\cite{DBLP:journals/corr/abs-2402-19173}}} & \multicolumn{1}{l|}{\multirow{3}{*}{}}                                                & n                 & 2024.2        & 7B            \\ \cline{3-5} 
\multicolumn{1}{l|}{}                                                                        & \multicolumn{1}{l|}{}                                                                 & n                 & 2024.2        & 15B           \\ \cline{3-5} 
\multicolumn{1}{l|}{}                                                                        & \multicolumn{1}{l|}{}                                                                 & y                 & 2024.4        & 15B           \\ \hline
\multicolumn{1}{l|}{WizardCoder\cite{luo2023wizardcoder}}                   & \multicolumn{1}{l|}{StarCoder}                                                                 & y                 & 2023.6        & 15B           \\ \hline
\multirow{6}{*}{CodeLlama\cite{DBLP:journals/corr/abs-2308-12950}}  & \multirow{6}{*}{Llama 2\cite{touvron2023llama}} & n                 & 2023.8        & 7B            \\ \cline{3-5} 
                                                                                     &                                                                  & y                 & 2023.8        & 7B            \\ \cline{3-5} 
                                                                                     &                                                                  & n                 & 2023.8        & 13B           \\ \cline{3-5} 
                                                                                     &                                                                  & y                 & 2023.8        & 13B           \\ \cline{3-5} 
                                                                                     &                                                                  & n                 & 2023.8        & 34B           \\ \cline{3-5} 
                                                                                     &                                                                  & y                 & 2023.8        & 34B           \\ \hline
\multirow{6}{*}{DeepSeekCoder\cite{deepseek-coder}}                 & \multirow{6}{*}{}                                                & n                 & 2023.10       & 1.3B          \\ \cline{3-5} 
                                                                                     &                                                                  & y                 & 2023.10       & 1.3B          \\ \cline{3-5} 
                                                                                     &                                                                  & n                 & 2023.10       & 6.7B          \\ \cline{3-5} 
                                                                                     &                                                                  & y                 & 2023.10       & 6.7B          \\ \cline{3-5} 
                                                                                     &                                                                  & n                 & 2023.10       & 33B           \\ \cline{3-5} 
                                                                                     &                                                                  & y                 & 2023.10       & 33B           \\ \hline

\multicolumn{1}{l|}{Magicoder\cite{DBLP:journals/corr/abs-2312-02120}}                              & \multicolumn{1}{l|}{DeepSeekCoder}                                                         & y                 & 2023.12       & 6.7B          \\ \hline
\end{tabular}
\end{table}
\subsection{Studied Code LLMs} \label{sec:baselines}
We select 19 state-of-the-art (SOTA) code LLMs which have been extensively examined in recent studies on code generation~\cite{luo2023wizardcoder, liu2023humanevalplus}. In particular, we focus on the open-source models released after 2023, while excluding smaller models (with fewer than 1 billion parameters) due to their limited efficacy and excluding larger models (with more than 20 billion parameters) due to computational resource constraints. Table~\ref{table:baseline_info} presents the details of the code LLMs studied in our experiments, including their release dates (Column ``Time''), the model sizes (Column ``Size''), the base model (Column ``Base Model''), and whether the model has been instruction-tuned (Column ``Instruct''). As shown in Table~\ref{table:baseline_info}, our study encompasses a diverse range of code LLMs, varying across multiple dimensions, such as (i) utilization of different base models, (ii) coverage of model sizes ranging from 1 billion to 15.5 billion parameters, (iii) presence or absence of instruction tuning, and (iv) inclusion of various versions of series models.

\subsection{Dataset}
\label{sec:datasets}
\begin{table}[]
\caption{Datasets Used for Running LLMs}
    \centering
\label{tab:dataset}
\begin{tabular}{lllll}
\hline
\multicolumn{2}{l}{DataSet}              & Size & Time                    & Language                \\ \hline
\multicolumn{2}{l}{HumanEval-P\cite{chen2021huamneval}}            & 164  & 2021.7                  & Python                  \\ \hline
\multicolumn{2}{l}{HumanEval-J\cite{athiwaratkun2023multilingual}} & 161  & 2022.10                 & Java                    \\ \hline
\multicolumn{2}{l}{MBPP\cite{austin2021mbpp}}                 & 974  & 2021.8                  & Python                  \\ \hline

\end{tabular}
\end{table}

We use three popular code generation datasets: HumanEval-Python, HumanEval-Java, and MBPP. HumanEval-Python and HumanEval-Java are the Python and Java versions of the HumanEval benchmark, featuring programming tasks with function signatures, docstrings, bodies, and unit tests. MBPP comprises about 1,000 crowd-sourced Python problems, each with a task description, solution, and three automated test cases. Statistical details of these datasets are shown in Table~\ref{tab:dataset}. Note that HumanEval-Python and MBPP provide solutions for each task, while HumanEval-Java does not.
\subsection{Metrics}\label{sec:metrics}
To quantitatively evaluate the repetition prevalence in an automated way, we leverage three metrics in RQ1, i.e., \textit{rep-n}, \textit{rep-line}, and \textit{sim-line}. In particular, while \textit{rep-n} is adapted from previous work on general text generation~\cite{DBLP:conf/nips/SuLWYKC22},  \textbf{rep-line} and \textbf{sim-line} are newly proposed in this study, which are designed to quantitatively evaluate the repetition of generated code. All the metrics range from 0 to 100, where higher values indicate higher repetition or similarity.

\parabf{Rep-n.} It measures the proportion of repeated n-grams in the generated code using Eq. \ref{eq:rep-line}, where \(\hat{x}\) represents the generated code. This metric calculates the percentage of non-unique n-grams, highlighting n-gram repetition. To compute \textbf{rep-n}, we split the code into words using a tokenizer. Unlike typical text tokenization, special characters (e.g., ``@'') except underscores (``\_'') are replaced with spaces, and the code is tokenized based on spaces. Regular expressions ensure complete identifiers are not split. For example, ``def min\_cost(cost, m, n):'' is tokenized into [``def'', ``min\_cost'', ``cost'', ``m'', ``n'']. The \textbf{rep-n} metric assesses redundancy by measuring repeated token sequences.

\begin{equation}
\label{eq:rep-n}
\textbf{rep-n} = 100 \times \left(1.0 - \frac{\left|\text{unique n-grams}(\hat{x})\right|}{\left|\text{total n-grams}(\hat{x})\right|}\right)
\end{equation}

\parabf{Rep-line.} It evaluates the proportion of repeated lines in the generated code using Eq. \ref{eq:rep-line}. This metric calculates the percentage of non-unique lines, considering lines identical if they are exact duplicates. Code is split by newline characters, with empty lines removed before calculation. The \textbf{rep-line} metric measures redundancy by assessing repeated lines in the code.

\begin{equation}
\label{eq:rep-line}
\textbf{rep-line} = 100 \times \left(1.0 - \frac{\left|\text{unique lines}(\hat{x})\right|}{\left|\text{total lines}(\hat{x})\right|}\right)
\end{equation}

\parabf{Sim-line.} It assesses line similarity in the generated code using edit distance (Levenshtein distance \cite{Levenshtein}). Lines are considered similar if their edit distance similarity exceeds 0.8. Defined by Eq. \ref{eq:sim-line}, the metric calculates the percentage of lines in dissimilar sets, highlighting diversity. Lines are obtained similarly to \textbf{rep-line}, with token-level edit distance calculated after tokenizing each line as for \textbf{rep-n}. The \textbf{sim-line} metric measures code diversity by evaluating line similarity.

\begin{equation}
\label{eq:sim-line}
\textbf{sim-line} = 100 \times \left(1.0 - \frac{\left|\text{dissimilar line sets}(\hat{x})\right|}{\left|\text{total lines}(\hat{x})\right|}\right)
\end{equation}

\subsection{Experimental Procedure}\label{sec:motivation:design}

\parabf{RQ1.}  For open-source LLMs, we use their released versions from official repositories, following the documentation. The maximum window length is set to 512 tokens, the smallest among the models studied. HumanEval-Java (H-J) and HumanEval-Python (H-P) provide prompts with function signatures and docstrings used directly as input. For MBPP, we extract function signatures and concatenate them with the task description, placing the description first. Models generate code based on these prompts. To explore code repetition and reduce randomness, we use greedy search. Experiments are conducted on eight A800-80G GPUs.

\parabf{RQ2 (Manual labeling).} 
To systematically identify and categorize repetition patterns in code generation, we sampled and analyzed LLM-generated code fragments with repetitive content. Each sample was inspected for recurring patterns. We also reviewed user feedback from LLM-based code completion tools, which highlighted common repetition issues in practice. Through iterative refinement, we defined and categorized repetition patterns into various levels, as shown in Table \ref{table:patterns}. This process continued until the categorization was stable and comprehensive.

\section{Empirical Results}

\begin{table*}[ht]
\caption{Repetition Metrics for Code LLMs on Datasets (H-P: HumanEval-Python, H-J: HumanEval-Java)}
\centering
\label{tab:motivation:result}
\scriptsize
\begin{tabular}{|l|llll|llll|llll|}
\hline
\multirow{2}{*}{}       & \multicolumn{4}{l|}{rep-3}                                                                              & \multicolumn{4}{l|}{rep-line}                                                            & \multicolumn{4}{l|}{sim-line}                                                            \\ \cline{2-13} 
                        & \multicolumn{1}{l|}{H-P} & \multicolumn{1}{l|}{H-J} & \multicolumn{1}{l|}{MBPP} & H-P + MBPP & \multicolumn{1}{l|}{H-P}    & \multicolumn{1}{l|}{H-J}    & \multicolumn{1}{l|}{MBPP}    & H-P + MBPP    & \multicolumn{1}{l|}{H-P}    & \multicolumn{1}{l|}{H-J}    & \multicolumn{1}{l|}{MBPP}    & H-P + MBPP    \\ \hline
    SantaCoder-1.1b              & \multicolumn{1}{l|}{49.8} & \multicolumn{1}{l|}{13.3} & \multicolumn{1}{l|}{39.6} & 41.1(+1144.1\%) & \multicolumn{1}{l|}{37.8} 
& \multicolumn{1}{l|}{10.8} & \multicolumn{1}{l|}{25.8} & 27.5(+5400.4\%) & \multicolumn{1}{l|}{64.8} & \multicolumn{1}{l|}{51.9} & \multicolumn{1}{l|}{53.5} 
& 55.1(+396.7\%)  \\ \hline
    StarCoder-15.5b               & \multicolumn{1}{l|}{39.7} & \multicolumn{1}{l|}{7.4} & \multicolumn{1}{l|}{10.8} & 15.0(+353.2\%) & \multicolumn{1}{l|}{30.8} & \multicolumn{1}{l|}{6.8} & \multicolumn{1}{l|}{8.0} & 11.3(+2161.7\%) & \multicolumn{1}{l|}{55.2} & \multicolumn{1}{l|}{48.3} & \multicolumn{1}{l|}{16.1} & 21.8(+96.0\%)  \\ \hline
    WizardCoder-15b-I             & \multicolumn{1}{l|}{7.3} & \multicolumn{1}{l|}{5.8} & \multicolumn{1}{l|}{3.9} & 4.4(+32.1\%) & \multicolumn{1}{l|}{4.1} & \multicolumn{1}{l|}{6.7} & \multicolumn{1}{l|}{0.8} & 1.3(+150.7\%) & \multicolumn{1}{l|}{16.9} & \multicolumn{1}{l|}{40.6} & \multicolumn{1}{l|}{6.6} & 8.1(-26.8\%)  \\ \hline
    Magicoder-6.7b-I               & \multicolumn{1}{l|}{8.7} & \multicolumn{1}{l|}{11.3} & \multicolumn{1}{l|}{4.7} & 5.3(+60.5\%) & \multicolumn{1}{l|}{2.8} & \multicolumn{1}{l|}{9.2} & \multicolumn{1}{l|}{1.3} & 1.5(+208.7\%) & \multicolumn{1}{l|}{18.3} & \multicolumn{1}{l|}{30.7} & \multicolumn{1}{l|}{12.5} & 13.4(+20.4\%)  \\ \hline
    StarCoder2-7b           & \multicolumn{1}{l|}{45.1} & \multicolumn{1}{l|}{22.8} & \multicolumn{1}{l|}{45.7} & 45.6(+1280.9\%) & \multicolumn{1}{l|}{37.1} & \multicolumn{1}{l|}{17.0} & \multicolumn{1}{l|}{32.7} & 33.4(+6571.6\%) & \multicolumn{1}{l|}{63.0} & \multicolumn{1}{l|}{55.5} & \multicolumn{1}{l|}{58.0} & 58.7(+429.2\%)  \\ \hline
    StarCoder2-15b          & \multicolumn{1}{l|}{43.1} & \multicolumn{1}{l|}{18.2} & \multicolumn{1}{l|}{41.2} & 41.4(+1155.8\%) & \multicolumn{1}{l|}{33.5} & \multicolumn{1}{l|}{20.7} & \multicolumn{1}{l|}{28.3} & 29.1(+5714.1\%) & \multicolumn{1}{l|}{57.6} & \multicolumn{1}{l|}{52.9} & \multicolumn{1}{l|}{51.1} & 52.1(+369.1\%)  \\ \hline
    StarCoder2-15b-I & \multicolumn{1}{l|}{38.9} & \multicolumn{1}{l|}{24.5} & \multicolumn{1}{l|}{45.2} & 44.2(+1240.7\%) & \multicolumn{1}{l|}{26.2} & \multicolumn{1}{l|}{18.5} & \multicolumn{1}{l|}{22.2} & 22.8(+4463.2\%) & \multicolumn{1}{l|}{50.6} & \multicolumn{1}{l|}{50.6} & \multicolumn{1}{l|}{47.9} & 48.3(+335.1\%)  \\ \hline
    CodeLlama-7b         & \multicolumn{1}{l|}{45.0} & \multicolumn{1}{l|}{14.7} & \multicolumn{1}{l|}{40.0} & 40.7(+1133.9\%) & \multicolumn{1}{l|}{34.2} & \multicolumn{1}{l|}{14.9} & \multicolumn{1}{l|}{27.1} & 28.1(+5521.6\%) & \multicolumn{1}{l|}{60.9} & \multicolumn{1}{l|}{45.7} & \multicolumn{1}{l|}{56.2} & 56.9(+412.2\%)            \\ \hline
CodeLlama-7b-I       & \multicolumn{1}{l|}{30.7} & \multicolumn{1}{l|}{11.0} & \multicolumn{1}{l|}{35.4} & 34.7(+952.6\%) & \multicolumn{1}{l|}{20.3} & \multicolumn{1}{l|}{11.6} & \multicolumn{1}{l|}{24.1} & 23.6(+4614.5\%) & \multicolumn{1}{l|}{47.2} & \multicolumn{1}{l|}{51.2} & \multicolumn{1}{l|}{50.5} & 50.1(+350.9\%)            \\ \hline
CodeLlama-13b        & \multicolumn{1}{l|}{43.1} & \multicolumn{1}{l|}{16.9} & \multicolumn{1}{l|}{47.9} & 47.2(+1329.8\%) & \multicolumn{1}{l|}{32.3} & \multicolumn{1}{l|}{16.6} & \multicolumn{1}{l|}{36.3} & 35.7(+7045.4\%) & \multicolumn{1}{l|}{58.5} & \multicolumn{1}{l|}{50.5} & \multicolumn{1}{l|}{61.4} & 61.0(+449.3\%)            \\ \hline
CodeLlama-13b-I      & \multicolumn{1}{l|}{23.7} & \multicolumn{1}{l|}{7.8} & \multicolumn{1}{l|}{23.9} & 23.8(+622.0\%) & \multicolumn{1}{l|}{17.6} & \multicolumn{1}{l|}{8.2} & \multicolumn{1}{l|}{16.7} & 16.8(+3263.8\%) & \multicolumn{1}{l|}{39.7} & \multicolumn{1}{l|}{47.8} & \multicolumn{1}{l|}{41.8} & 41.5(+273.6\%)            \\ \hline
CodeLlama-34b        & \multicolumn{1}{l|}{25.5} & \multicolumn{1}{l|}{8.0} & \multicolumn{1}{l|}{15.9} & 17.3(+423.6\%) & \multicolumn{1}{l|}{18.9} & \multicolumn{1}{l|}{11.2} & \multicolumn{1}{l|}{6.3} & 8.1(+1514.1\%) & \multicolumn{1}{l|}{46.6} & \multicolumn{1}{l|}{47.0} & \multicolumn{1}{l|}{32.8} & 34.8(+213.5\%)            \\ \hline
CodeLlama-34b-I      & \multicolumn{1}{l|}{25.0} & \multicolumn{1}{l|}{9.6} & \multicolumn{1}{l|}{19.7} & 20.5(+521.0\%) & \multicolumn{1}{l|}{16.6} & \multicolumn{1}{l|}{11.6} & \multicolumn{1}{l|}{9.5} & 10.5(+2005.0\%) & \multicolumn{1}{l|}{44.6} & \multicolumn{1}{l|}{48.7} & \multicolumn{1}{l|}{38.0} & 38.9(+250.7\%)            \\ \hline
DeepSeekCoder-1.3b   & \multicolumn{1}{l|}{31.8} & \multicolumn{1}{l|}{11.3} & \multicolumn{1}{l|}{11.5} & 14.4(+336.1\%) & \multicolumn{1}{l|}{25.5} & \multicolumn{1}{l|}{9.8} & \multicolumn{1}{l|}{6.0} & 8.8(+1658.1\%) & \multicolumn{1}{l|}{46.9} & \multicolumn{1}{l|}{39.8} & \multicolumn{1}{l|}{22.9} & 26.4(+137.6\%)            \\ \hline
DeepSeekCoder-1.3b-I & \multicolumn{1}{l|}{35.5} & \multicolumn{1}{l|}{13.2} & \multicolumn{1}{l|}{34.8} & 34.9(+958.3\%) & \multicolumn{1}{l|}{16.3} & \multicolumn{1}{l|}{8.7} & \multicolumn{1}{l|}{17.4} & 17.2(+3348.4\%) & \multicolumn{1}{l|}{38.3} & \multicolumn{1}{l|}{38.8} & \multicolumn{1}{l|}{37.2} & 37.4(+236.6\%)            \\ \hline
DeepSeekCoder-6.7b   & \multicolumn{1}{l|}{38.4} & \multicolumn{1}{l|}{9.7} & \multicolumn{1}{l|}{7.5} & 12.0(+263.1\%) & \multicolumn{1}{l|}{31.1} & \multicolumn{1}{l|}{8.8} & \multicolumn{1}{l|}{3.7} & 7.7(+1430.9\%) & \multicolumn{1}{l|}{52.0} & \multicolumn{1}{l|}{35.3} & \multicolumn{1}{l|}{19.0} & 23.8(+114.4\%)            \\ \hline
DeepSeekCoder-6.7b-I & \multicolumn{1}{l|}{34.3} & \multicolumn{1}{l|}{19.5} & \multicolumn{1}{l|}{29.0} & 29.8(+802.3\%) & \multicolumn{1}{l|}{21.5} & \multicolumn{1}{l|}{18.2} & \multicolumn{1}{l|}{24.1} & 23.7(+4643.8\%) & \multicolumn{1}{l|}{44.9} & \multicolumn{1}{l|}{46.8} & \multicolumn{1}{l|}{48.5} & 48.0(+332.6\%)            \\ \hline
DeepSeekCoder-33b    & \multicolumn{1}{l|}{29.3} & \multicolumn{1}{l|}{8.5} & \multicolumn{1}{l|}{8.5} & 11.5(+248.9\%) & \multicolumn{1}{l|}{23.9} & \multicolumn{1}{l|}{8.0} & \multicolumn{1}{l|}{5.0} & 7.8(+1453.6\%) & \multicolumn{1}{l|}{45.1} & \multicolumn{1}{l|}{33.4} & \multicolumn{1}{l|}{18.4} & 22.2(+100.3\%)            \\ \hline
DeepSeekCoder-33b-I  & \multicolumn{1}{l|}{30.7} & \multicolumn{1}{l|}{19.8} & \multicolumn{1}{l|}{36.3} & 35.5(+975.2\%) & \multicolumn{1}{l|}{23.3} & \multicolumn{1}{l|}{20.3} & \multicolumn{1}{l|}{23.7} & 23.6(+4623.1\%) & \multicolumn{1}{l|}{43.5} & \multicolumn{1}{l|}{51.5} & \multicolumn{1}{l|}{46.8} & 46.3(+317.3\%)            \\ \hline
Ground Truth  & \multicolumn{1}{l|}{3.8} & \multicolumn{1}{l|}{-} & \multicolumn{1}{l|}{3.2} & 3.3 & \multicolumn{1}{l|}{0.7} & \multicolumn{1}{l|}{-} & \multicolumn{1}{l|}{0.5} & 0.5 & \multicolumn{1}{l|}{15.4} & \multicolumn{1}{l|}{-} & \multicolumn{1}{l|}{10.3} & 11.1  \\ \hline
\end{tabular}
\end{table*}
\begin{figure}[htb]
	\centering
        \vspace{-1mm}
	\includegraphics[width=1.0\columnwidth]{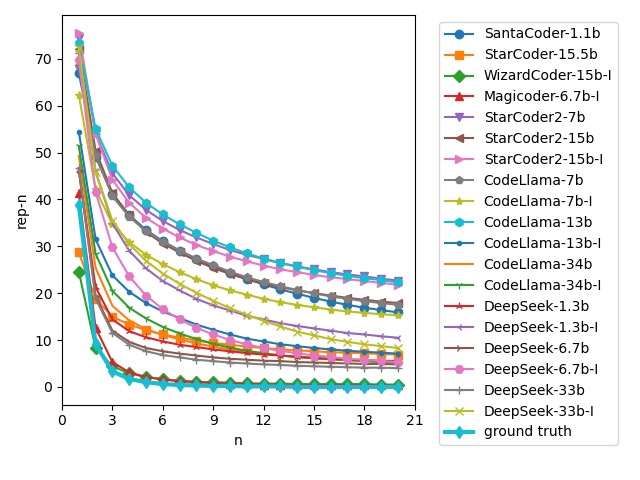}
	\caption{rep\_n across Different n-grams on H-P+MBPP}
	\label{fig:rep-n:distribution}
        \vspace{-1mm}
\end{figure}
\begin{figure}[htb]
	\centering
        \vspace{-1mm}	\includegraphics[width=1.0\columnwidth]{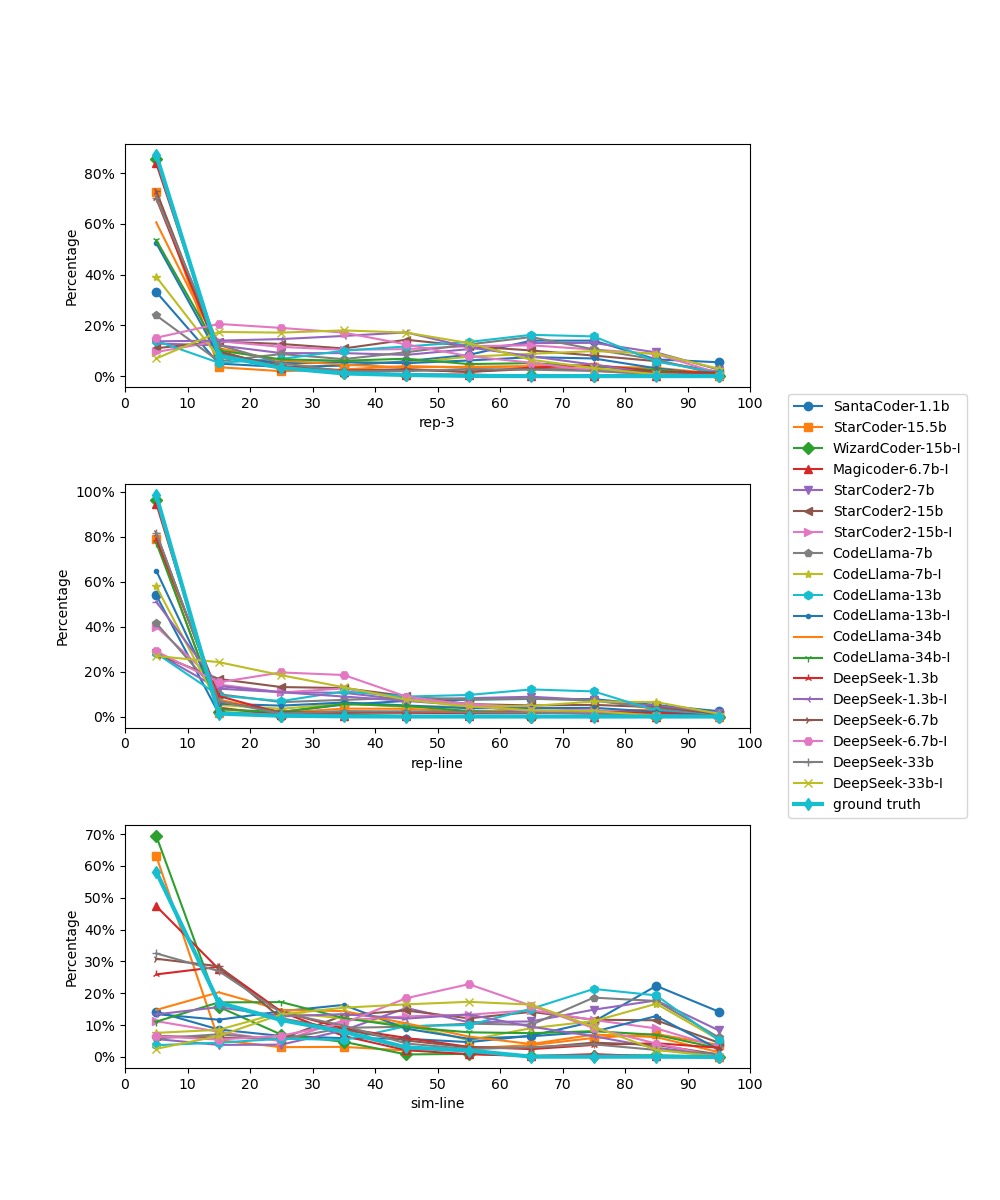}
	\caption{Distribution of rep-3, rep-line, and sim-line Metrics across Models}
        \label{fig:metrics:distribution}
        \vspace{-1mm}
\end{figure}

\subsection{RQ1 Results}
\label{sec:motivation:result}
Table~\ref{tab:motivation:result}, Figure~\ref{fig:rep-n:distribution}, and Figure~\ref{fig:metrics:distribution} present the three repetition metrics of studied  code LLMs. \textbf{Based on the results, we can have the overall finding that repetition issues are very prevalent across different code LLMs and different programming languages.} We then analyze the results from multiple perspectives. 

\subsubsection{RQ1.a (Repetition via different Metrics)}
Figure~\ref{fig:metrics:distribution} illustrates the distribution of repetition metrics (rep-3, rep-line, and sim-line) for various models on the H-P+MBPP dataset. Each plot reveals how different models handle repetition in code generation compared to the ground truth.

\parabf{Rep-n Analysis.}
The top plot displays the rep-3 values, which measure the repetition of three-gram sequences. The ground truth shows almost no repetition, indicating that human-written code rarely includes repetitive three-gram sequences. In contrast, most models demonstrate significant rep-3 values, particularly at the lower percentage end, which indicates a high level of short-sequence repetition. This highlights a common issue in generated code where models tend to reproduce similar three-word sequences multiple times.

Figure~\ref{fig:rep-n:distribution} shows the rep-n values on the H-P+MBPP dataset across different n-gram sizes. As the size of the n-gram increases, the ground truth exhibits minimal repetition, approaching zero. In contrast, most models still exhibit significant repetition, indicating that repetition in generated code is not limited to small-granularity n-grams but extends to larger chunks of code. This includes not only longer sequences of statements but even blocks of multiple statements.

\parabf{Rep-line Analysis.}
The middle plot shows the rep-line values, which track the repetition of entire lines of code. Similar to the rep-3 metric, the ground truth maintains very low rep-line values, indicating minimal repetition of whole lines in human-written code. However, most models still exhibit notable rep-line values, especially at lower percentages. This suggests that these models often generate redundant lines of code, leading to less diverse and potentially bloated outputs.

\parabf{Sim-line Analysis.}
The bottom plot presents the sim-line values, reflecting the similarity between lines of code. While the ground truth remains close to zero, showing little to no similarity between lines, the models display varying degrees of sim-line values. This indicates that generated code not only repeats exact lines but also produces lines that are highly similar to one another. Such similarity can result in code that is functionally redundant or overly verbose.

The sim-line metric indicates how similar the lines are within the generated code. Higher values suggest more repetitive patterns.
Models like StarCoder2-15b and CodeLlama-13b exhibit high sim-line values for Python tasks, indicating that their generated lines are more similar to each other. For example, StarCoder2-15b has a sim-line value of 57.6\% for H-P, which is high. Instruction-tuned models tend to have lower sim-line values, indicating more diverse line generation. For instance, DeepSeekCoder-6.7b-I has a sim-line value of 44.9\% for H-P, compared to its non-tuned version which has 52.0\%.

Overall, the significant differences between the ground truth and model-generated code across all three metrics indicate a pervasive issue with repetition in automated code generation. While human-written code maintains a high level of diversity and minimal redundancy, models struggle to replicate this quality. Instead, they often produce repetitive and similar sequences, lines, and patterns.

\subsubsection{RQ1.b (Repetition across Different Models)}
Overall, as evidenced by high values in rep-3 and rep-line metrics, most LLMs exhibit substantial repetition in their generated code. In contrast, the human-written code (\ie{} the repetition metrics in the ground-truth row) has much less repetition compared to the code generated by LLMs. First, WizardCoder-15b-I and Magicoder-6.7b-I show the low repetition metrics among all the studied models. For example,  WizardCoder-15b-I shows the lowest repetition metrics  (i.e., rep-3 value of 4.4 for H-P + MBPP, and a rep-line value of 1.3) across all categories, making it one of the best performers; Magicoder-6.7b-I performs with low repetition rates, particularly noticeable in the H-P + MBPP dataset with a rep-line value of 1.5\%. Second, SantaCoder-1.1b and StarCoder2-7b show much higher repetition metrics among all the studied models. For example, SantaCoder-1.1b shos  high repetition across all metrics, particularly in Python tasks. Its rep-3 value for H-P is 49.8\%; while StarCoder2-7b shows high repetition rates with a rep-3 value of 45.1\% for H-P. 

\parabf{Impact of Model Size.} Larger model sizes tend to alleviate repetition issues. For exampke, larger models such as StarCoder-15.5b and StarCoder2-15b generally show better performance with lower repetition rates compared to smaller models like SantaCoder-1.1b and DeepSeekCoder-1.3b. For instance, StarCoder-15.5b has a rep-3 value of 15.0\% for H-P + MBPP, whereas SantaCoder-1.1b has 41.1\%. 

\parabf{Impact of Instruction Tuning.} Models with instruction tuning generally mitigate repetition problems more effectively. For example, models with instruction tuning (denoted by ``-I'') tend to perform better in reducing repetition. For example, WizardCoder-15b-I and Magicoder-6.7b-I show significantly lower rep-3 and rep-line values compared to their non-instruction-tuned counterparts. WizardCoder-15b-I has a rep-3 value of 4.4\% for H-P + MBPP, which is much lower than many other models.

\begin{figure}[htb]
	\centering
        \vspace{-1mm}\includegraphics[width=0.8\columnwidth]{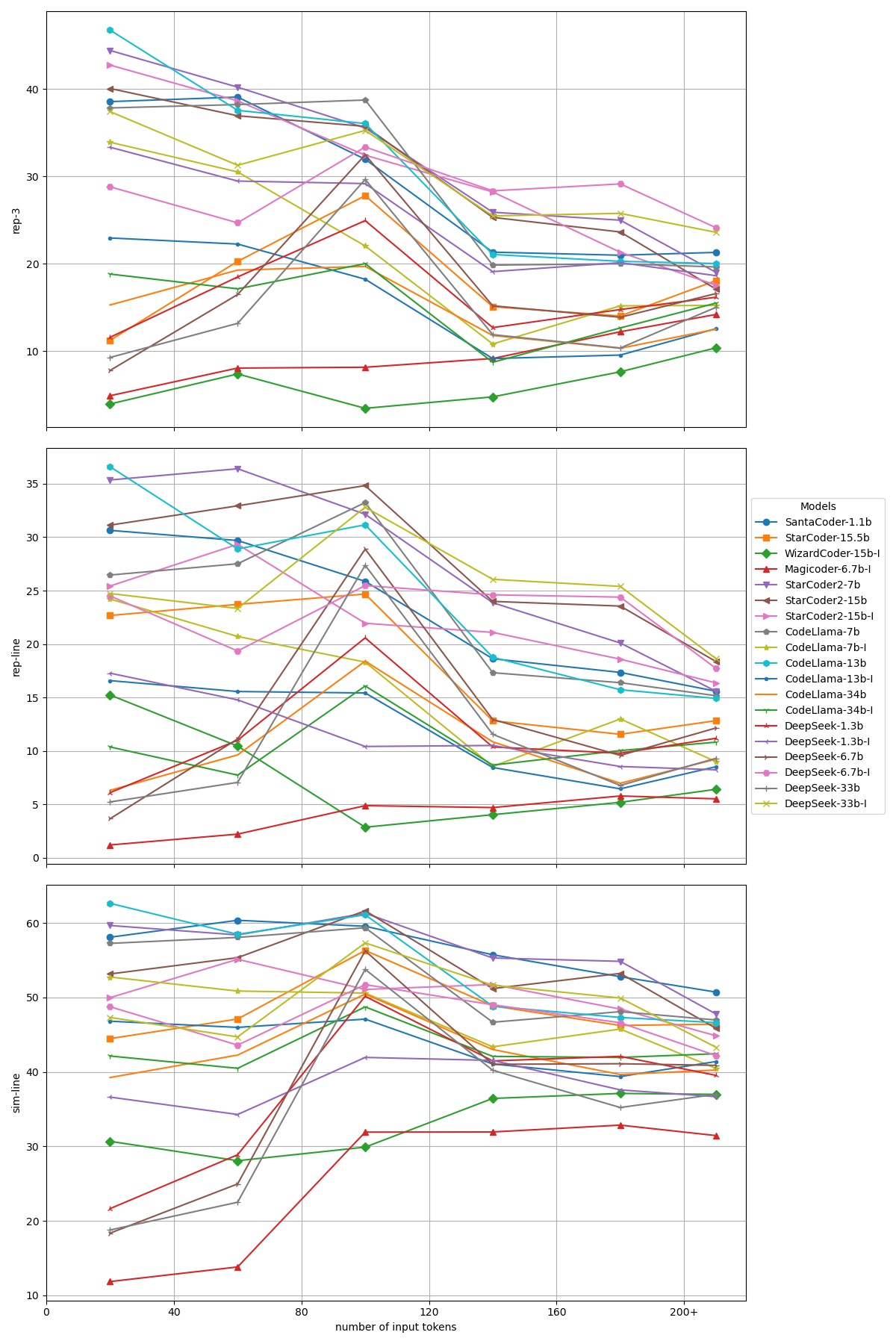}
	\caption{Trends of Metrics by Input Tokens Number Intervals}
        \label{fig:impact:input}	
        \vspace{-1mm}
\end{figure}
\begin{figure}[htb]
	\centering
        \vspace{-1mm}\includegraphics[width=0.8\columnwidth]{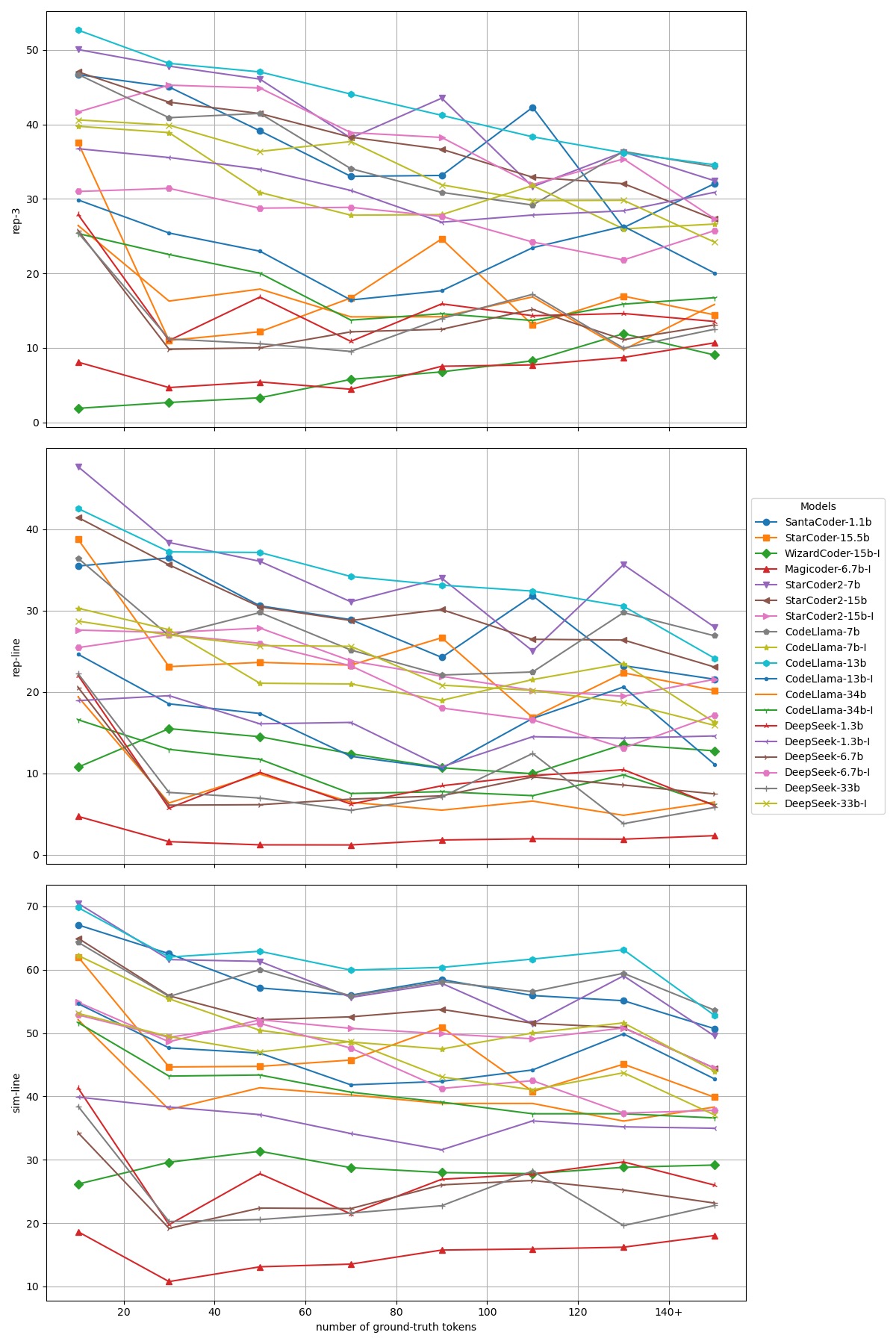}
	\caption{Trends of Repetition Metrics by Ground Truth Tokens Number}
        \label{fig:impact:output}	
        \vspace{-1mm}
\end{figure}

\subsubsection{RQ1.c (Input/Output Impact on repetition)}
We further analyze how the input/output can impact the repetition issues in code generation. 

\parabf{Input Impact.}
Figure~\ref{fig:impact:input} presents the trends of three repetition metrics (rep-n, rep-line, and sim-line) by the number of input tokens, e.g., the prompt consisting of the function signature and the docstrings.  

\parabf{Output Impact.}
Figure~\ref{fig:impact:output} shows how three repetition metrics in code generation vary with the number of ground truth tokens, which indicates task difficulty. The horizontal axis represents ground truth tokens (20 to 140+), and the vertical axis shows the metric values. Each point reflects the average metric value for an LLM across tasks with ground truth tokens in a given range.

\textbf{Rep-3 Trend Analysis.}
The first subplot shows that as the number of ground truth tokens increases, rep-n values generally decrease, suggesting that more complex tasks result in less repeated n-grams. Larger models (e.g., 33b) exhibit lower rep-n values compared to smaller models (e.g., 1.1b and 3b), indicating better performance in generating varied n-grams even with increasing task difficulty.

\textbf{Rep-Line Trend Analysis.}
The second subplot illustrates the proportion of repeated lines in the generated code. As task difficulty increases, rep-line values generally decrease, suggesting that more complex tasks result in fewer repeated lines. Larger models tend to perform better, showing lower rep-line values compared to smaller models.

\begin{table*}[ht]
\scriptsize
\centering
\caption{Repetition Patterns and Examples}
\label{table:patterns}
\begin{tabular}{|l|l|l|l|l|}
\hline
\textbf{Granularity}       & \textbf{Repetitive Content}               & \textbf{Example}                                                                                                                               & \textbf{Repetitive Content}                      & \textbf{Example}                                                                                                                                                                                                        \\ \hline
\multirow{4}{*}{Character} & Numeric Literal                           & min\_count = 9999999999...                                                                                                                     & Identifier                                       & map\_size\_reverse\_size\_reverse\_...                                                                                                                                                                                  \\ \cline{2-5} 
                           & String Literal                            & print(count\_occurance('stdstds...                                                                                                             & Conditional Statement                            & return 1 if a[0] == b[0] and a[1] == b[1] and...                                                                                                                                                                        \\ \cline{2-5} 
                           & \makecell[l]{Dictionary Key-Value\\Pairs} & \makecell[l]{assert ascii\_hash("abc") == \{97: 32, 98: 33,\\ 99: 34, 100: 35, 101: 36...}                                                     & Chained Function Calls                           & \makecell[l]{return s.replace("!?", "?").replace("??", "?")\\.replace("!!", "?")... }                                                                                                                                                   \\ \cline{2-5} 
                           & Array Elements                            & nums = [1,2,3,4,5,6,7,8,9...                                                                                                                   &                                                  &                                                                                                                                                                                                                         \\ \hline
\multirow{4}{*}{Statement} & Test Statements                           & \makecell[l]{assert count(0) == 1\\assert count(1) == 1}                                                                                       & \makecell[l]{Chained Attribute\\Accesses}        & \makecell[l]{root.right.left = TreeNode(6)\\root.right.right = TreeNode(7)\\root.right.left.left = TreeNode(8)}                                                                                                         \\ \cline{2-5} 
                           & Assignment Statements                     & \makecell[l]{int i = 0;\\int j = 0;\\int k = 0;}                                                                                               & \makecell[l]{Dictionary Key-Value\\Pairs}        & \makecell[l]{\{\\"1":1,\\"2":2,}                                                                                                                                                                                        \\ \cline{2-5} 
                           & Comments                                  & \makecell[l]{\# 8. If the list is not empty, \\return the length of the list\\\# 9. If the list is not empty, \\return the length of the list} & Array Elements                                   & \makecell[l]{[\\1, 2, 3,\\1, 2, 3,\\1, 2, 3,}                                                                                                                                                                           \\ \cline{2-5} 
                           & Empty Lines                                & \textbackslash n\textbackslash n\textbackslash n...                                                                                            &                                                  &                                                                                                                                                                                                                         \\ \hline
\multirow{3}{*}{Block}     & Functions                                 & \makecell[l]{def make\_changeamount():\\...\\def make\_changeamount():\\...}                                                                   & \makecell[l]{Comment + Assignment \\ Statements} & \makecell[l]{\# Input1\\ s = "xabb"\\ \# Input2\\ s = "xabb"}                                                                                                                                                           \\ \cline{2-5} 
                           & Comments                                  & \makecell[l]{\# Example 1:\\...\\\# Example 2:\\...}                                                                                           & \makecell[l]{Comment + Test \\ Statements}       & \makecell[l]{\# Test case 1\\ supw\_time = [3, 2, 1, 1, 2, 3, 1, 3, 2, 1]\\ assert min\_supw\_time(supw\_time) == 4\\ \# Test case 2\\ supw\_time = [3, 2, 3, 2, 3, 5, 1, 3]\\ assert min\_supw\_time(supw\_time) == 5} \\ \cline{2-5} 
                           & Conditional Statements                    & \makecell[l]{if k == 2:\\return maxarr - minarr\\if k == 3:\\return maxarr - minarr}                                                           & Special Characters                               & \makecell[l]{\# \#\\ \# \#\\ \#\# \# \#\#\\ \#\#\#\#\#\#\#}                                                                                                                                                             \\ \hline
\end{tabular}
\end{table*}

\subsection{RQ2 Results (Repetition Patterns)}

Table~\ref{table:patterns} shows the 20 repetition patterns we identified across three granularities: character, statement, and block.

\begin{itemize}[leftmargin=15pt]
    \item Character Level (7 patterns): This level includes repetitions within individual elements in a statement, such as numeric and string literals, dictionary key-value pairs, array elements, identifiers, conditional statements, and chained function calls.
    \item Statement Level (7 patterns): This level involves repetitions of entire lines of code such as test statements, assignment statements, comments, array elements, chained attribute accesses, empty lines, and dictionary key-value pairs.
    \item Block Level (6 patterns): This level encompasses repetitions of larger code structures like functions, comments, conditional statements, assignment statements combined with comments, test statements combined with comments, and special character art.
\end{itemize}

In summary, repetition patterns vary by level: character-level involves small units within lines, statement-level covers entire lines, and block-level affects larger code structures. Recognizing these patterns aids in devising strategies to minimize repetition and enhance code quality.

\parabf{Repetition Extent.}
In code generation, LLMs often produce repetitive content up to the maximum length. Repetition patterns can be categorized by their nature and extent:

\begin{itemize}[leftmargin=15pt]
    \item \textbf{Complete Repetition}: The model generates the exact same content repeatedly until it hits the maximum length. For example, repeatedly generating ``x = 1'' in a loop.
    \item \textbf{Similar Repetition}: The model generates content with a consistent pattern but slight variations, such as test assertions with different parameters like ``assertEqual(func(1), 2)'' and ``assertEqual(func(2), 3)''.
    \item \textbf{Finite Repetition}: Repetition occurs a specific number of times and then stops, often seen with character-level repetitions within a line. An example is creating an array like ``[1, 1, 1, 2, 2]''.
    \item \textbf{Infinite Repetition}: The model continues generating repetitive content indefinitely until the maximum length is reached, such as repeating the same function definition.
    \item \textbf{Random Repetition}: The model generates random sequences up to the maximum length, like a long string of random numbers such as ``3.1415926...''.
\end{itemize}

These categories describe various degrees of repetition in LLM-generated code, providing insights into how and why repetition occurs. Detailed examples are available in our replication package \cite{decoder2024}.

\section{Rule-based Repetition Mitigation}
\label{sec:app:migration}
Building on our empirical findings, we propose \ourtool{}, a lightweight rule-based approach to detect and repair repetitive patterns in LLM-generated code. The system operates in two phases:  (1) \textbf{detection} of repetitive code segments (Section~\ref{sec:app_detection}), and (2) \textbf{repair} of the repetitive parts in the generated code through pruning (Section~\ref{sec:app_repair}).

\subsection{Repetition Detection}
\label{sec:app_detection}
Our repetition detection algorithm first determines whether a code snippet contains repetition. If repetition is present, it then identifies its scope and classifies it into one of several predefined patterns (see Table~\ref{table:patterns}). 

The overall process is a pipeline, formalized in Algorithm~\ref{alg:rep_dectect_pipeline}. Given an input code snippet, the algorithm outputs: (1) the granularity of the repetition, (2) the pattern type (as defined in Table~\ref{table:patterns}), and (3) the repeated units, including their content and the start and end positions of all occurrences. If no repetition is detected, both the granularity and pattern type are returned as None. Notably, partially repeated segments at the end of the code—those that are incomplete due to truncation—are still considered valid repeated units. 

The pipeline follows a cascading strategy, progressing from fine-grained to coarse-grained patterns (character-level → statement-level → block-level), which balances computational efficiency with the practical frequency of each pattern type observed in LLM-generated code. We split the code into lines based on newline characters to facilitate different level  repetition detection.

We next detail three repetition detection algorithms at different levels of granularity: character-level (Algorithm~\ref{alg:char-rep-detect}), statement-level (Algorithm~\ref{alg:state-rep-detect}), and block-level (Algorithm~\ref{alg:block-rep-detect}). Each algorithm shares the same input and output format.

\parabf{Character-level Repetition Detection: Algorithm~\ref{alg:char-rep-detect}}

Character-level repetition often manifests as excessive code generation within a single line, resulting in duplicated characters, tokens, or identifiers. As shown in Algorithm~\ref{alg:char-rep-detect}, the detection process begins by examining the last generated line. If this line exceeds a predefined length threshold (e.g., 150 characters) or contains end-of-sequence markers such as ``<|endoftext|>'', it is considered highly indicative of repetition. In such cases, the algorithm attempts to extract repeated segments directly from this line using pattern-specific rules (e.g., detecting repeated numeric literals or punctuation sequences).

If no repetition is found in the last line, the algorithm proceeds to scan the remaining lines in reverse order. For each line, it applies a set of syntactic and semantic rules defined for character-level patterns in Table~\ref{table:patterns}. These include patterns such as Dictionary Key-Value Pair Repetition, where repeated key-value structures are identified. The scan halts as soon as a valid repetition pattern is detected, returning the pattern type along with the corresponding repeated units, each annotated with its content and positional information.

For example, the line ``my\_dict = \{"a": 1, "a": 2, "a": 3\}'' will be detected as a Dictionary Key-Value Pair Repetition, with all occurrences of ``"a": 1''  extracted as repeated units.

\parabf{Statement-level Repetition Detection: Algorithm~\ref{alg:state-rep-detect}}

Statement-level repetition refers to repeated or highly similar lines occurring consecutively within the code. To detect such patterns, Algorithm~\ref{alg:state-rep-detect} performs a sequential scan of code lines, identifying the longest contiguous block of repeated statements, including those with minor variations.

During the scan, each line is compared with its preceding line using a similarity function. If their similarity exceeds a predefined threshold, the line is marked as part of the current repetition block. The algorithm tracks the longest such block by maintaining its start position and length to ensure comprehensive detection.

Line-level similarity is computed using TF-IDF vectorization and cosine similarity~\cite{TFIDF}, with a threshold set to 0.65 based on empirical tuning. This threshold can be adjusted to suit different levels of repetition tolerance. To handle truncation scenarios—where a large model generates excessive repetition but the output ends abruptly—the last line is checked using a prefix-based similarity heuristic to determine whether it belongs to an ongoing repetition unit.
Once the longest contiguous repetition block is located, the algorithm invokes \textsc{ExtractStatementRepUnits} to extract the repeated units (i.e., line content and occurrence positions) and classify them according to predefined statement-level repetition patterns, as outlined in Table~\ref{table:patterns}.

\parabf{Block-level Repetition Detection: Algorithm~\ref{alg:block-rep-detect}}

Block-level repetition involves the appearance of identical or highly similar multi-line sequences in a contiguous manner. Based on the observation that such repetitive units typically exhibit consistent line counts, our detection algorithm performs an iterative search over different block lengths to identify repeated segments. Specifically, it assumes block lengths ranging from a minimum length $L_{\min}$ (default 2) up to a maximum length $L_{\max}$, which we empirically set as $2n/3$, where $n$ is the total number of lines. This threshold strikes a balance between detection coverage and computational efficiency.

For each assumed block length $L$, the algorithm applies a sliding window strategy over the code lines: at each position $i$, it compares the current block $B_1 = lines[i : i + L - 1]$ with the next block $B_2 = lines[i + L : i + 2L - 1]$. If these two blocks are sufficiently similar, they are considered as repeated units. The block similarity is computed using the \textsc{IsSimilar} function, which is also used in statement-level detection. This function applies TF-IDF vectorization and cosine similarity with a threshold of 0.65. The algorithm then uses \textsc{FindAllRepeats} to extend the repetition detection forward and collect all matching blocks. 
If the number of detected units exceeds the previous best, it updates the current best result.
To address cases of incomplete code due to truncation, the algorithm includes a prefix similarity check to determine whether the last partial block belongs to a repetitive unit. Once the best repeated region is identified, the repetition is further classified into specific block-level patterns based on predefined rules. 

Static analysis tools like Tree-sitter are utilized to support the parsing of syntactic structures such as statements and identifiers. Our implementation supports multiple widely-used programming languages, including Java, Go, JavaScript/TypeScript, Python, and C++.

\begin{algorithm}[ht]
\footnotesize
\captionsetup{font=footnotesize} 
\caption{Repetition Detection Pipeline}
\label{alg:rep_dectect_pipeline}
\begin{algorithmic}[1]
\Require 
    Code snippet $C$
\Ensure 
    Granularity $G \in \{\textsc{CHARACTER}, \textsc{Statement}, \textsc{Block}\}$,
    Pattern $P \in \mathbb{S}$,
    Units $\mathcal{U} = [(s_1, e_1, \text{content}_1), \dots]$

\State $lines \gets \textsc{SplitLines}(C)$

\State $(P, \mathcal{U}) \gets \textsc{IntraLineDetection}(lines)$
\If{$P \neq \textsc{None}$} \Return $(\textsc{CHARACTER}, P, \mathcal{U})$ \EndIf

\State $(P, \mathcal{U}) \gets \textsc{SingleLineDetection}(lines)$
\If{$P \neq \textsc{None}$} \Return $(\textsc{Statement}, P, \mathcal{U})$ \EndIf

\State $(P, \mathcal{U}) \gets \textsc{MultiLineDetection}(lines)$
\If{$P \neq \textsc{None}$} \Return $(\textsc{Block}, P, \mathcal{U})$ \EndIf

\State \Return $(None, None, [])$
\end{algorithmic}
\end{algorithm}

\begin{algorithm}[ht]
\footnotesize
\captionsetup{font=footnotesize} 
\caption{Character-level Repetition Detection}
\label{alg:char-rep-detect}
\begin{algorithmic}[1]
\Require 
    Code lines $lines[1..n]$
\Ensure 
    Pattern $P \in \mathbb{S}$,
    Units $\mathcal{U} = [(s_1, e_1, \text{content}_1), \dots]$
\Function{DetectCharRep}{$lines$}
    \State $n \gets |lines|$
    \If{$n > 0$ and $\textsc{CheckLineOverlength}(lines[n])$}
        \State   $(P, \mathcal{U}) \gets \textsc{ExtractCharRepetitionUnits}(lines[n])$
        \If{$P \neq \textsc{None}$}
        \State   \Return $(P, \mathcal{U})$
        \EndIf
    \EndIf
    
    \For{$i \gets n$ \textbf{downto} $1$ } 
        \State $(P, \mathcal{U}) \gets \textsc{ExtractCharRepetitionUnits}(lines[i])$
        \If{$P \neq \textsc{None}$}
        \State   \Return $(P, \mathcal{U})$
        \EndIf
    \EndFor
    
    \State \Return $(\textsc{None}, [])$ 
\EndFunction
\end{algorithmic}
\end{algorithm}

\begin{algorithm}[ht]
\footnotesize
\captionsetup{font=footnotesize} 
\caption{Statement-level Repetition Detection}
\label{alg:state-rep-detect}
\begin{algorithmic}[1]
\Require 
    Code lines $lines[1..n]$
\Ensure 
    Pattern $P \in \mathbb{S}$,
    Units $\mathcal{U} = [(s_1, e_1, \text{content}_1), \dots]$
\Function{DetectStatementRep}{$lines$}
\State $max\_len \gets 0$, $max\_start \gets -1$, $current\_len \gets 1$
    
    \For{$i \gets 2$ to $n$}
        \If{\textsc{IsSimilar}$(lines[i], lines[i-1])$}
            \State $current\_len \gets current\_len + 1$
            \If{$current\_len > max\_len$}
                \State $max\_len \gets current\_len$
                \State $max\_start \gets i - current\_len + 1$
            \EndIf
        \Else
            \State $current\_len \gets 1$ \Comment{Reset counter}
        \EndIf
    \EndFor

     \If{$max\_len \geq 2$} \Comment{Minimum 2 consecutive lines}
        \State $s \gets max\_start$, $e \gets s + max\_len - 1$
        \State $(P, \mathcal{U}) \gets \textsc{ExtractStatementRepUnits}(lines[s..e])$
        \State \Return $(P, \mathcal{U})$
    \EndIf
    
    \State \Return $(\textsc{None}, [])$
\EndFunction
\end{algorithmic}
\end{algorithm}

\begin{algorithm}[ht]
\footnotesize
\captionsetup{font=footnotesize} 
\caption{Block-level Repetition Detection}
\label{alg:block-rep-detect}
\begin{algorithmic}[1]
\Require 
    Code lines $lines[1..n]$
\Ensure 
    Pattern $P \in \mathbb{S}$,
    Units $\mathcal{U} = [(s_1, e_1, \text{content}_1), \dots]$
\Function{DetectBlockRep}{$lines$} 
    \State $n \gets |lines|$, $best\_units \gets []$, $best\_pattern  \gets \textsc{None}$
    
    \For{$L \gets L_{\min}$ to $L_{\max}$} \Comment{Check all block sizes}
        \For{$i \gets 0$ to $n - 2L$} \Comment{sliding window for repetition blcoks}
            \State $B_1 \gets lines[i:i+L-1]$ \Comment{Reference block}
            \State $B_2 \gets lines[i+L:i+2L-1]$ \Comment{Next candidate block}
            
            \If{$\textsc{IsSimilar}(B_1, B_2)$}
                \State $\mathcal{U} \gets \textsc{FindAllRepeats}(lines, i, L)$
                \If{$|\mathcal{U}| > |best\_units|$}
                    \State $best\_units \gets \mathcal{U}$
                \EndIf
            \EndIf
        \EndFor
    \EndFor
    
    \If{$best\_units \neq []$}
        \State $best\_pattern \gets \textsc{IdentifyBlockPattern}(best\_units)$
        \State \Return $(best\_pattern, best\_units)$
    \Else
        \State \Return $(\textsc{None}, [])$
    \EndIf
\EndFunction
\end{algorithmic}
\end{algorithm}

\subsection{Repetition Repair}

Building on the results of our repetition detection algorithm, we precisely locate each instance of repeated units within the generated code. Based on these positions, we implement a lightweight repair mechanism designed to eliminate redundant repetitions while preserving the surrounding semantic structure.

The core strategy is straightforward: for each group of consecutive repeated units, we retain only the first valid occurrence and remove all subsequent duplicates. This operation is applied consistently across both statement-level and block-level repetition cases. For instance, in a pattern such as \texttt{A; B; B; B; C; D}, our repair method identifies the repeated segment \texttt{B; B; B} and simplifies it to a single instance, resulting in \texttt{A; B; C; D}.

\begin{figure}[htb]
	\centering
        \vspace{-1mm}
	\includegraphics[width=1\columnwidth]{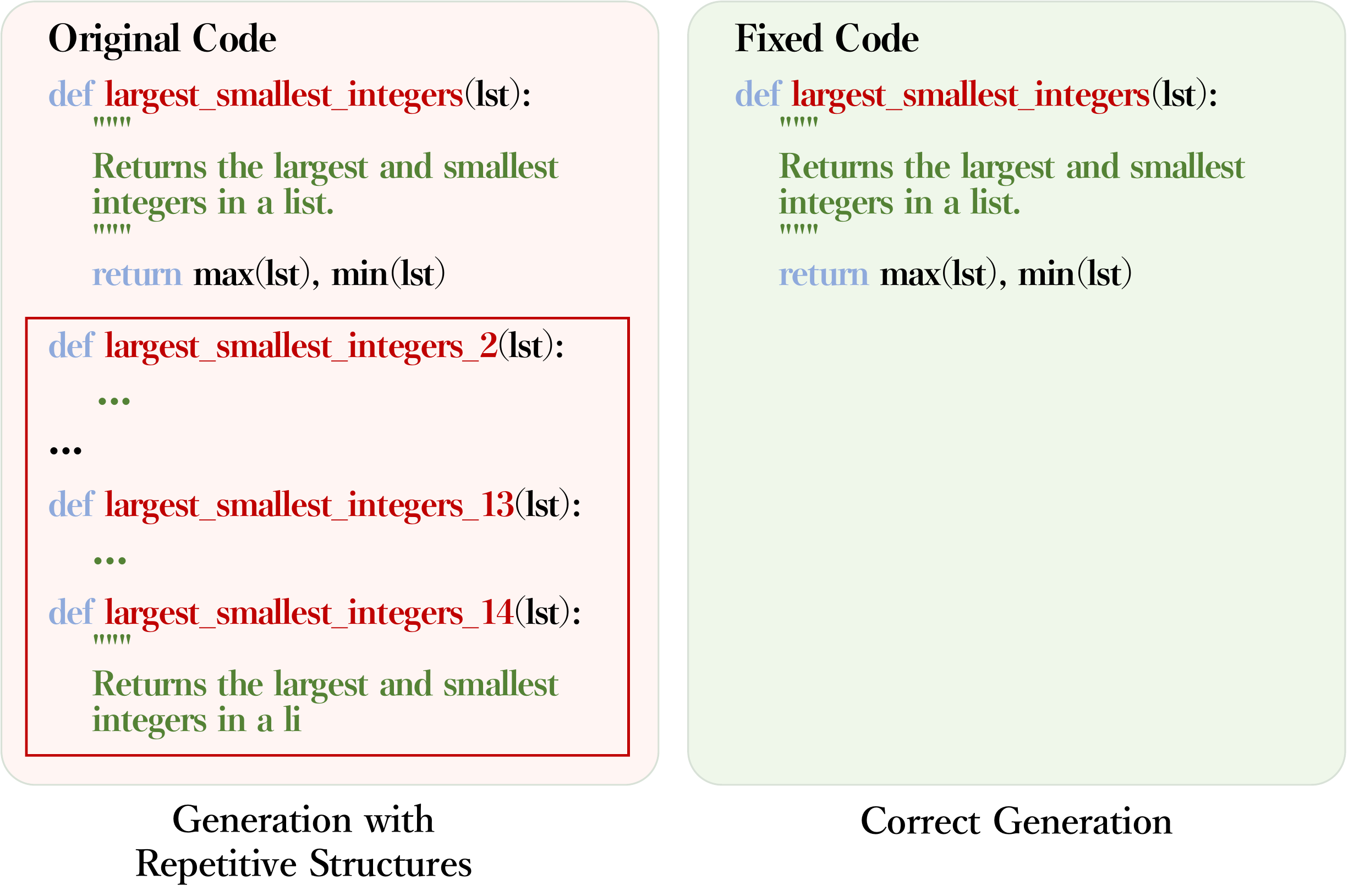}
	\caption{Examples of Code Repair Using Our Approach}
	\label{fig:fixed code example}
        \vspace{-1mm}
\end{figure}

As illustrated in Figure~\ref{fig:fixed code example}, in the code with function repetitions, the first function \texttt{largest\_smallest\_integers(lst)} is identified as the initial valid repetition unit. The red-boxed area highlights the subsequent identified repetition units, which are highly similar to the first instance and are removed. After the repair process, we preserve the first valid function definition and eliminate all subsequent duplicate function definitions, including the last incomplete one.

\label{sec:app_repair}

\section{Evaluation}
\label{sec:eval}
In this section, we evaluate the effectiveness and performance of our proposed \ourtool technique. We address the following RQs to comprehensively assess our approach:

\begin{itemize}[leftmargin=15pt]
    \item \textbf{RQ3: (Repair Effectiveness) How effective is DeRep in repairing detected repetitions?}
    \begin{itemize}
    \item \textbf{RQ3.a} How does DeRep perform in comparison to baseline methods?
    \item \textbf{RQ3.b} How does DeRep contribute to enhancing existing methods?
    \end{itemize}
    \item \textbf{RQ4: (Detected Pattern Distribution) What is the distribution of detected repetition patterns in the generated code? }
    \item \textbf{RQ5: (Industry Setting Performance) How does DeRep perform in an industry setting?}
\end{itemize}

\subsection{RQ3 (Repair Effectiveness)}
\label{sec:eval:rq3}

\parabf{Baselines.}  
we compare DeRep with several general repetition mitigation techniques as follows:
\begin{itemize}[leftmargin=15pt]
    \item \textbf{Beam Search} \cite{freitag2017beam}:
    Beam search generates multiple sequences at each step and keeps the top \( k \) sequences based on their cumulative probabilities. This method aims to find the most likely sequence of tokens by exploring multiple potential paths and selecting the best one.
    
    \item \textbf{Top-p Sampling} \cite{DBLP:conf/iclr/nsample}: Given a probability distribution over words, it selects the smallest subset of words whose cumulative probability exceeds a threshold \( p \), and then samples from this subset to generate the next word. It aims to balance diversity and relevance by restricting token generation to a subset of high-probability options.
    
    \item \textbf{Top-k Sampling} \cite{DBLP:conf/iclr/nsample}: In this method, tokens are sampled from the top \( k \) most probable candidates at each time step. This approach reduces repetitiveness by focusing on a fixed number of likely tokens.
    
    \item \textbf{Contrastive Search} \cite{DBLP:conf/nips/SuLWYKC22}:
    Contrastive search selects tokens from the most probable candidates while ensuring that each generated token is sufficiently distinct from the preceding context. This method maintains semantic coherence and avoids generating repetitive or degenerate outputs.
    
    \item \textbf{Repetition Penalty} \cite{DBLP:journals/corr/abs-1909-05858}:
    This technique applies a penalty to the probability distribution of tokens that have already appeared in the generated sequence, discouraging the reuse of previously generated tokens and mitigating repetitive patterns.

\end{itemize}

\parabf{Setup and Metrics.}
For evaluation, we use the HumanEval-Python and MBPP datasets as described in Section \ref{sec:datasets}. To comprehensively evaluate the repetition in the generated code, we define a new metric, rep, which is the average of three existing metrics: rep-n, rep-line, and sim-line. Additionally, we compute the Pass@1 metric to evaluate whether a method can effectively reduce code repetition without compromising the correctness. For both our approach and the baselines, we use DeepSeekCoder as the backbone LLM, with greedy search results. We also explore various hyperparameter settings: for beam search, we tried beam sizes of 3, 5, and 10; for top-k sampling, we used \( k = 3, 5, 10 \); for top-p sampling, we set \( p = 0.95, 0.9, 0.85 \); for contrastive search, we used a similarity threshold of 0.6; and for repetition penalty, we applied values of 1.2, 1.5, and 2.0.

\subsubsection{RQ3.a (Comparison)}
Table \ref{tab-rq3-pre} compares \app{} with baselins in terms of repair effectiveness and generation correctness. Due to space limit, only the optimal parameter thresholds for each method are included. DeepSeekCoder is abbreviated as DSC for brevity.

\textbf{DeRep Effectiveness.} Overall, \app{} significantly outperforms existing repetition mitigation techniques by effectively eliminating redundant code while improving code generation accuracy. Applying our DeRep method to greedy search yields average improvement of 88.3\% in rep metric. Although slightly less effective than Repetition Penalty in reducing repetition, DeRep notably enhances Pass@1 scores, showing an average increase of 208.3\% over greedy search results, achieving an average score of 33.3. Notably, for smaller models such as DeepSeekCoder-1.3b-I, DeRep achieves a Pass@1 score of 40.4\%, surpassing all general methods and even outperforming the larger DeepSeekCoder-33b-I model. This demonstrates the robustness of DeRep in enhancing both repetition reduction and functionality correctness.
\begin{tcolorbox}[colback=myfinding!50, colframe=white, width=\linewidth, arc=3mm, boxrule=0.5mm, left=2mm, right=2mm, top=2mm, bottom=2mm, boxsep=0mm]
\textbf{Finding 1}: \app{} significant outperformexisting repetition mitigation techniques, effectively eliminating redundant code while enhancing code generation accuracy.
\end{tcolorbox}
\textbf{Repetition Mitigation Techniques Analysis.} 
Various general repetition mitigation techniques achieve improvements over the native greedy search results. For rep, reductions range from 25.7\% to 94.5\%. Among these techniques, Repetition Penalty with a threshold of 1.2 stands out as the most effective, achieving the lowest average scores of 1.7 for rep, with reductions of 94.5\%. However, despite improvements in repetition reduction, these techniques compromise functionality correctness. For instance, Repetition Penalty-1.2 significantly reduces rep from 29.1 to 0.5 for DeepSeekCoder-1.3b-I but also decreases Pass@1 from 18.5 to 0.2.
\begin{tcolorbox}[colback=myfinding!50, colframe=white, width=\linewidth, arc=3mm, boxrule=0.5mm, left=2mm, right=2mm, top=2mm, bottom=2mm, boxsep=0mm]
\textbf{Finding 2}: While Repetition Penalty outperforms other baselines in reducing repetition, it exhibits a poor trade-off between repetition reduction and functionality correctness.
\end{tcolorbox}

\subsubsection{RQ3.b (Enhancement)}
As DeRep is orthogonal to existing baseline methods, it can be integrated with any of them to enhance their effectiveness. To demonstrate the potential improvements \app{} can bring, we applied DeRep to the outputs of five baseline methods. Table~\ref{tab-rq3-post} presents the performance results after applying \app{} to these baselines. 

\textbf{Integrating DeRep with General Repetition Mitigation Techniques.} 
When integrated with \app{}, existing repetition mitigation techniques exhibit substantial improvements across all models. The average improvements range from 87.9\%-97.0\% for rep metric. Additionally, Pass@1 scores exhibit average improvements of 53.7\%-215.7\%. These results demonstrate the scalability of DeRep and its effectiveness in enhancing performance across various models.
\begin{tcolorbox}[colback=myfinding!50, colframe=white, width=\linewidth, arc=3mm, boxrule=0.5mm, left=2mm, right=2mm, top=2mm, bottom=2mm, boxsep=0mm]
\textbf{Finding 3}: \app{} can enhance general-purpose repetition mitigation methods by performing effective post-processing on their generated outputs to further reduce repeated code.
\end{tcolorbox}

\begin{table}[]
\centering
\caption{Performance Comparison of Repetition Mitigation Techniques (CT: Contrastive Search)}
\label{tab-rq3-pre}
\scriptsize
\setlength{\tabcolsep}{0pt}
\begin{tabular}{|llllllll|}
\hline
\multicolumn{1}{|l|}{}           & \multicolumn{1}{c|}{Greedy} & \multicolumn{1}{c|}{Beam(3)} & \multicolumn{1}{c|}{Top-k(10)} & \multicolumn{1}{c|}{Top-p(0.85)} & \multicolumn{1}{c|}{CT(0.6)}          & \multicolumn{1}{l|}{Penalty(1.2)}     & DeRep(Greedy) \\ \hline
\multicolumn{8}{|c|}{rep}                                                                                                                                                                                                                                       \\ \hline
\multicolumn{1}{|l|}{DSC-1.3b}   & \multicolumn{1}{l|}{34.9}   & \multicolumn{1}{l|}{21.9}    & \multicolumn{1}{l|}{5.9}       & \multicolumn{1}{l|}{8.9}         & \multicolumn{1}{l|}{1.1}              & \multicolumn{1}{l|}{0.6}              & 4.7           \\ \hline
\multicolumn{1}{|l|}{DSC-1.3b-I} & \multicolumn{1}{l|}{29.1}   & \multicolumn{1}{l|}{25.4}    & \multicolumn{1}{l|}{12.9}      & \multicolumn{1}{l|}{17.3}        & \multicolumn{1}{l|}{15.3}             & \multicolumn{1}{l|}{0.5}              & 3.7           \\ \hline
\multicolumn{1}{|l|}{DSC-6.7b}   & \multicolumn{1}{l|}{36.5}   & \multicolumn{1}{l|}{23.2}    & \multicolumn{1}{l|}{6.0}       & \multicolumn{1}{l|}{8.8}         & \multicolumn{1}{l|}{2.3}              & \multicolumn{1}{l|}{1.0}              & 4.4           \\ \hline
\multicolumn{1}{|l|}{DSC-6.7b-I} & \multicolumn{1}{l|}{31.1}   & \multicolumn{1}{l|}{27.2}    & \multicolumn{1}{l|}{12.7}      & \multicolumn{1}{l|}{16.1}        & \multicolumn{1}{l|}{11.4}             & \multicolumn{1}{l|}{3.4}              & 3.1           \\ \hline
\multicolumn{1}{|l|}{DSC-33b}    & \multicolumn{1}{l|}{27.0}   & \multicolumn{1}{l|}{15.9}    & \multicolumn{1}{l|}{5.8}       & \multicolumn{1}{l|}{10.0}        & \multicolumn{1}{l|}{3.8}              & \multicolumn{1}{l|}{1.2}              & 3.0           \\ \hline
\multicolumn{1}{|l|}{DSC-33b-I}  & \multicolumn{1}{l|}{25.9}   & \multicolumn{1}{l|}{23.3}    & \multicolumn{1}{l|}{7.1}       & \multicolumn{1}{l|}{11.3}        & \multicolumn{1}{l|}{6.2}              & \multicolumn{1}{l|}{3.1}              & 2.5           \\ \hline
\multicolumn{1}{|l|}{Average}    & \multicolumn{1}{l|}{30.7}   & \multicolumn{1}{l|}{22.8}    & \multicolumn{1}{l|}{8.5}       & \multicolumn{1}{l|}{12.1}        & \multicolumn{1}{l|}{6.7}              & \multicolumn{1}{l|}{1.7}              & 3.6           \\ \hline
\multicolumn{1}{|l|}{Pre-DeRep}  & \multicolumn{1}{c|}{-}      & \multicolumn{1}{c|}{-25.7\%} & \multicolumn{1}{c|}{-72.3\%}   & \multicolumn{1}{c|}{-60.6\%}     & \multicolumn{1}{c|}{-78.2\%}          & \multicolumn{1}{c|}{\textbf{-94.5\%}} & -88.3\%       \\ \hline
\multicolumn{8}{|c|}{Pass@1}                                                                                                                                                                                                                                      \\ \hline
\multicolumn{1}{|l|}{DSC-1.3b}   & \multicolumn{1}{l|}{2.9}    & \multicolumn{1}{l|}{11.0}    & \multicolumn{1}{l|}{13.2}      & \multicolumn{1}{l|}{11.3}        & \multicolumn{1}{l|}{11.0}             & \multicolumn{1}{l|}{7.7}              & 22.6          \\ \hline
\multicolumn{1}{|l|}{DSC-1.3b-I} & \multicolumn{1}{l|}{18.5}   & \multicolumn{1}{l|}{9.4}     & \multicolumn{1}{l|}{12.9}      & \multicolumn{1}{l|}{7.6}         & \multicolumn{1}{l|}{8.1}              & \multicolumn{1}{l|}{0.2}              & 40.4          \\ \hline
\multicolumn{1}{|l|}{DSC-6.7b}   & \multicolumn{1}{l|}{5.7}    & \multicolumn{1}{l|}{17.7}    & \multicolumn{1}{l|}{15.5}      & \multicolumn{1}{l|}{12.8}        & \multicolumn{1}{l|}{17.7}             & \multicolumn{1}{l|}{10.6}             & 24.9          \\ \hline
\multicolumn{1}{|l|}{DSC-6.7b-I} & \multicolumn{1}{l|}{8.2}    & \multicolumn{1}{l|}{8.2}     & \multicolumn{1}{l|}{12.1}      & \multicolumn{1}{l|}{9.5}         & \multicolumn{1}{l|}{23.1}             & \multicolumn{1}{l|}{17.0}             & 43.5          \\ \hline
\multicolumn{1}{|l|}{DSC-33b}    & \multicolumn{1}{l|}{14.4}   & \multicolumn{1}{l|}{19.3}    & \multicolumn{1}{l|}{14.0}      & \multicolumn{1}{l|}{16.9}        & \multicolumn{1}{l|}{21.2}             & \multicolumn{1}{l|}{22.7}             & 30.0          \\ \hline
\multicolumn{1}{|l|}{DSC-33b-I}  & \multicolumn{1}{l|}{15.3}   & \multicolumn{1}{l|}{24.8}    & \multicolumn{1}{l|}{22.4}      & \multicolumn{1}{l|}{18.0}        & \multicolumn{1}{l|}{29.5}             & \multicolumn{1}{l|}{31.3}             & 38.5          \\ \hline
\multicolumn{1}{|l|}{Average}    & \multicolumn{1}{l|}{10.8}   & \multicolumn{1}{l|}{15.1}    & \multicolumn{1}{l|}{15.0}      & \multicolumn{1}{l|}{12.7}        & \multicolumn{1}{l|}{18.4}             & \multicolumn{1}{l|}{14.9}             & 33.3          \\ \hline
\multicolumn{1}{|l|}{Pre-DeRep}  & \multicolumn{1}{c|}{-}      & \multicolumn{1}{c|}{+39.8\%} & \multicolumn{1}{c|}{+38.9\%}   & \multicolumn{1}{c|}{+17.6\%}     & \multicolumn{1}{c|}{\textbf{+70.4\%}} & \multicolumn{1}{c|}{+38.0\%}          & +208.3\%      \\ \hline

\end{tabular}
\end{table}

\begin{table}[]
\centering
\caption{Performance of Combining DeRep and General Repetition Mitigation Techniques (CT: Contrastive Search)}
\label{tab-rq3-post}
\scriptsize
\setlength{\tabcolsep}{0pt}
\begin{tabular}{|llllll|}
\hline
\multicolumn{1}{|l|}{}           & \multicolumn{1}{c|}{Beam(3)}           & \multicolumn{1}{c|}{Top-k(10)} & \multicolumn{1}{c|}{Top-p(0.85)} & \multicolumn{1}{c|}{CT(0.6)}  & Penalty(1.2)                          \\ \hline
\multicolumn{6}{|c|}{rep}                                                                                                                                                                                             \\ \hline
\multicolumn{1}{|l|}{DSC-1.3b}   & \multicolumn{1}{l|}{4.9}               & \multicolumn{1}{l|}{3.7}       & \multicolumn{1}{l|}{3.3}         & \multicolumn{1}{l|}{1.0}      & 0.6                                   \\ \hline
\multicolumn{1}{|l|}{DSC-1.3b-I} & \multicolumn{1}{l|}{3.5}               & \multicolumn{1}{l|}{3.5}       & \multicolumn{1}{l|}{3.7}         & \multicolumn{1}{l|}{3.2}      & 0.4                                   \\ \hline
\multicolumn{1}{|l|}{DSC-6.7b}   & \multicolumn{1}{l|}{4.4}               & \multicolumn{1}{l|}{3.1}       & \multicolumn{1}{l|}{3.5}         & \multicolumn{1}{l|}{1.8}      & 1.0                                   \\ \hline
\multicolumn{1}{|l|}{DSC-6.7b-I} & \multicolumn{1}{l|}{3.5}               & \multicolumn{1}{l|}{2.9}       & \multicolumn{1}{l|}{3.3}         & \multicolumn{1}{l|}{2.8}      & 1.5                                   \\ \hline
\multicolumn{1}{|l|}{DSC-33b}    & \multicolumn{1}{l|}{2.8}               & \multicolumn{1}{l|}{2.3}       & \multicolumn{1}{l|}{2.8}         & \multicolumn{1}{l|}{1.6}      & 0.9                                   \\ \hline
\multicolumn{1}{|l|}{DSC-33b-I}  & \multicolumn{1}{l|}{3.2}               & \multicolumn{1}{l|}{2.3}       & \multicolumn{1}{l|}{2.2}         & \multicolumn{1}{l|}{1.6}      & 1.3                                   \\ \hline
\multicolumn{1}{|l|}{Average}    & \multicolumn{1}{l|}{3.7}               & \multicolumn{1}{l|}{3.0}       & \multicolumn{1}{l|}{3.1}         & \multicolumn{1}{l|}{2.0}      & 0.9                                   \\ \hline
\multicolumn{1}{|l|}{Post-DeRep} & \multicolumn{1}{c|}{-87.9\%}           & \multicolumn{1}{c|}{-90.3\%}   & \multicolumn{1}{c|}{-89.8\%}     & \multicolumn{1}{c|}{-93.5\%}  & \multicolumn{1}{c|}{\textbf{-97.0\%}} \\ \hline
\multicolumn{6}{|c|}{Pass@1}                                                                                                                                                                                          \\ \hline
\multicolumn{1}{|l|}{DSC-1.3b}   & \multicolumn{1}{l|}{23.5}              & \multicolumn{1}{l|}{13.2}      & \multicolumn{1}{l|}{16.1}        & \multicolumn{1}{l|}{11.3}     & 7.7                                   \\ \hline
\multicolumn{1}{|l|}{DSC-1.3b-I} & \multicolumn{1}{l|}{38.1}              & \multicolumn{1}{l|}{31.4}      & \multicolumn{1}{l|}{33.6}        & \multicolumn{1}{l|}{28.4}     & 0.4                                   \\ \hline
\multicolumn{1}{|l|}{DSC-6.7b}   & \multicolumn{1}{l|}{27.5}              & \multicolumn{1}{l|}{16.4}      & \multicolumn{1}{l|}{20.8}        & \multicolumn{1}{l|}{17.3}     & 11.1                                  \\ \hline
\multicolumn{1}{|l|}{DSC-6.7b-I} & \multicolumn{1}{l|}{44.7}              & \multicolumn{1}{l|}{38.8}      & \multicolumn{1}{l|}{42.6}        & \multicolumn{1}{l|}{40.0}     & 26.3                                  \\ \hline
\multicolumn{1}{|l|}{DSC-33b}    & \multicolumn{1}{l|}{28.5}              & \multicolumn{1}{l|}{21.2}      & \multicolumn{1}{l|}{22.7}        & \multicolumn{1}{l|}{26.5}     & 22.2                                  \\ \hline
\multicolumn{1}{|l|}{DSC-33b-I}  & \multicolumn{1}{l|}{42.0}              & \multicolumn{1}{l|}{27.8}      & \multicolumn{1}{l|}{34.9}        & \multicolumn{1}{l|}{34.3}     & 31.9                                  \\ \hline
\multicolumn{1}{|l|}{Average}    & \multicolumn{1}{l|}{34.1}              & \multicolumn{1}{l|}{24.8}      & \multicolumn{1}{l|}{28.5}        & \multicolumn{1}{l|}{26.3}     & 16.6                                  \\ \hline
\multicolumn{1}{|l|}{Post-DeRep} & \multicolumn{1}{c|}{\textbf{+215.7\%}} & \multicolumn{1}{c|}{+129.6\%}  & \multicolumn{1}{c|}{+163.9\%}    & \multicolumn{1}{c|}{+143.5\%} & \multicolumn{1}{c|}{+53.7\%}          \\ \hline

\end{tabular}
\end{table}

\subsection{RQ4 (Detected Pattern Distribution)}
\label{sec:eval:rq3}
\begin{figure*}[htb]
	\centering
        \vspace{-1mm}	\includegraphics[width=2.0\columnwidth]{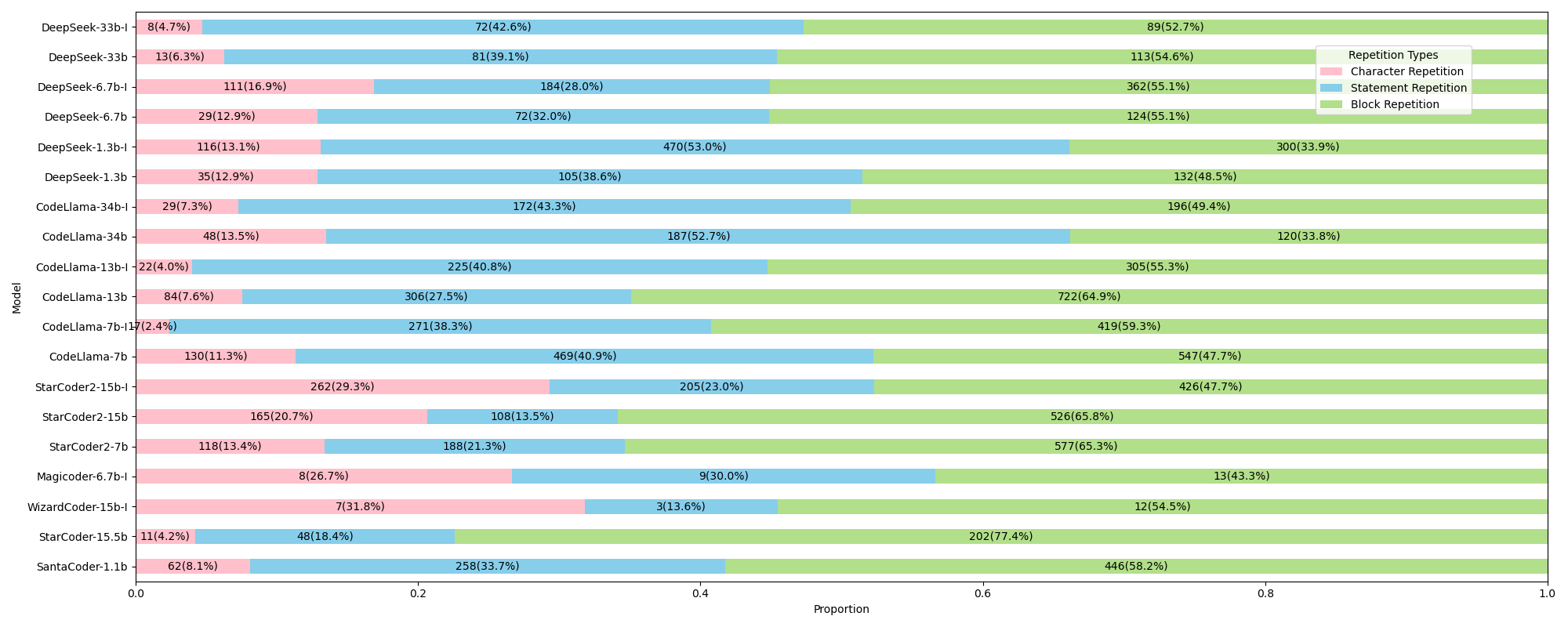}
	\caption{Repetition Pattern Proportions Across Different Code LLMs}
        \label{fig:proportions:repetition pattern}
        \vspace{-1mm}
\end{figure*}
\begin{figure}[htb]
	\centering
        \vspace{-1mm}	\includegraphics[width=1.0\columnwidth]{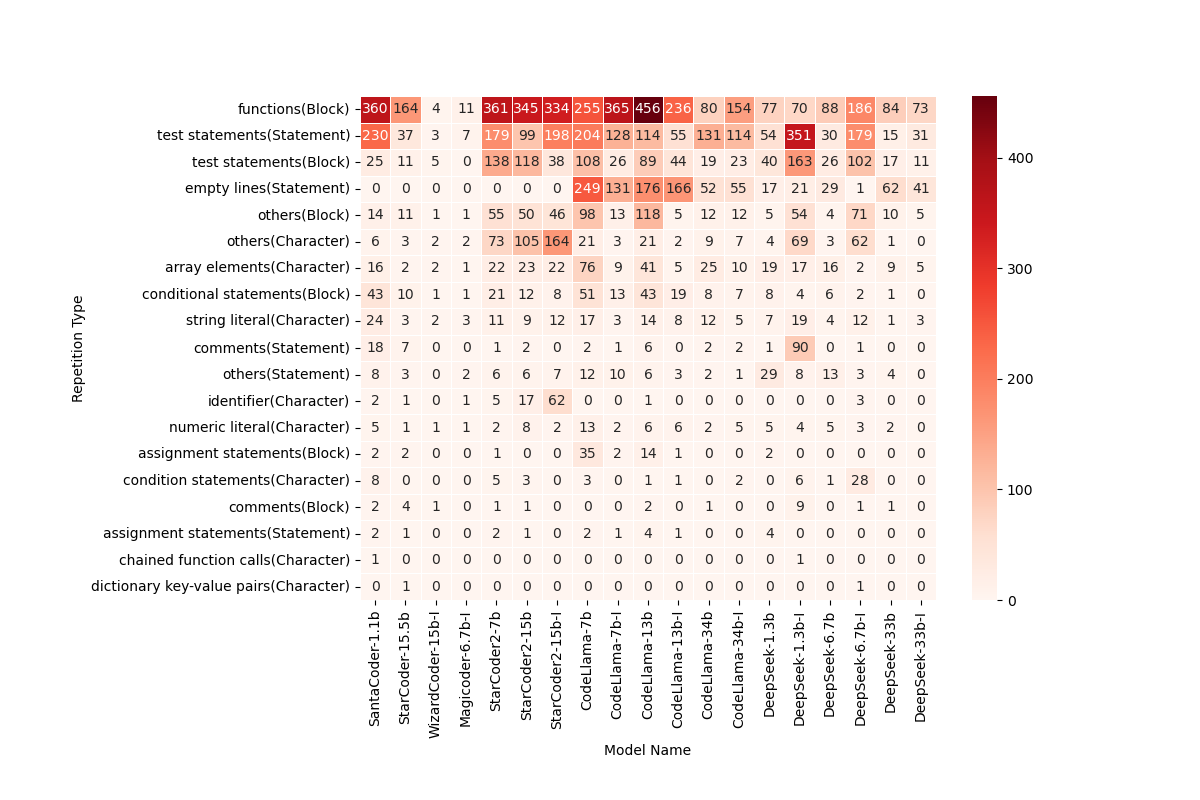}
	\caption{Repetition Pattern per LLM}
        \label{fig:repetition pattern: LLM}
        \vspace{-1mm}
\end{figure}

In this RQ, we analyze the repetition patterns in code generated by various LLMs using our developed repetition detection algorithm. 
Our study focuses on identifying and categorizing different types of repetitions to understand their prevalence and distribution across different models.

Figure \ref{fig:proportions:repetition pattern} presents the proportion of different repetition types (character, statement, and block) across various LLMs. The data shows that block-level repetitions are predominant in most models, with models like DeepSeek-6.7b-I and CodeLlama-13b showing the highest proportions. This suggests a common issue where models tend to replicate larger code structures rather than individual statements or characters. Statement-level repetitions are also significant, with models like StarCoder-2.7b and StarCoder2-15b-I displaying higher proportions, indicating that while these models are better at varying blocks, they still struggle with statement-level diversity. Character-level repetitions, although less frequent overall, are noticeable in models like StarCoder-15.5b and SantaCoder-1.1b, suggesting that finer granularity variations are better managed by most models, though some still exhibit repetition at this level.

Figure \ref{fig:repetition pattern: LLM} provides a detailed heatmap showing the frequency of various repetition types across different LLMs. From Figure \ref{fig:repetition pattern: LLM}, we observe that function blocks exhibit the highest repetition frequencies, particularly in models like StarCoder-2.7b, WizardCoder-15b-I, and StarCoder-1.1b. This indicates a common tendency for these models to replicate entire function blocks, suggesting challenges in generating unique functional structures. Similarly, test statements, both at the statement and block levels, show significant repetition in models such as StarCoder-2.7b and WizardCoder-15b-I. This highlights potential issues in generating diverse test cases. Furthermore, array elements and conditional statements also display noticeable repetition, particularly in models like CodeLlama-13b and StarCoder-2.7b. Repetitions of string literals and identifiers, while less frequent, are still present, especially in models like WizardCoder-15b-I and CodeLlama-7b, indicating some difficulties in varying these finer elements.

The results from these figures indicate that while LLMs have made significant strides in generating coherent code, there are still notable challenges related to repetition. Function blocks and test statements, in particular, are prone to high levels of repetition.

\subsection{RQ5 (Industry Setting Performance)}
\label{sec:eval:rq5}

Our tool, DeRep, has been integrated into a code completion tool within our partner company. It is currently used in production to post-process code completion results generated by large models. Feedback from over 50 internal users, collected through sampling interviews, indicates a significant reduction in code repetition and an enhanced user experience.

To further understand the effectiveness of our method on industrial data, we conducted a sampling analysis during a one-week trial period across multiple pilot development departments. We collected 5,000 code completion results in six programming languages, respectively: Java, Go, JavaScript/TypeScript, Python, and C++. These results were then processed using our DeRep method, and changes in rep-3, rep-line, and sim-line metrics were recorded. Table \ref{tab-industrial-case} presents the detailed results, demonstrating a reduction in repetition metrics across all programming languages, which underscores the effectiveness of our DeRep method.

One notable advantage of our method is its speed, providing detection and repair results at the millisecond level, making it suitable for real-time applications and integration into existing code completion tools. Our experiments showed that the average detection time per code snippet is around 50ms.

In summary, DeRep not only effectively reduces code repetition but also meets the real-time performance requirements for industrial applications, enhancing the overall code completion experience.

\begin{table*}[]
\centering
\caption{Changes in Industrial Case Metrics Pre- and Post-DeRep Application}
\label{tab-industrial-case}
\footnotesize
\begin{tabular}{|l|lll|lll|lll|}
\hline
\multirow{2}{*}{} & \multicolumn{3}{l|}{rep-3}                                         & \multicolumn{3}{l|}{rep-line}                                      & \multicolumn{3}{l|}{sim-line}                                      \\ \cline{2-10} 
                  & \multicolumn{1}{l|}{Pre-DeRep} & \multicolumn{1}{l|}{Post-DeRep} & improve & \multicolumn{1}{l|}{Pre-DeRep} & \multicolumn{1}{l|}{Post-DeRep} & improve & \multicolumn{1}{l|}{Pre-DeRep} & \multicolumn{1}{l|}{Post-DeRep} & improve \\ \hline
Java              & \multicolumn{1}{l|}{21.5}       & \multicolumn{1}{l|}{19.9}      &    -7.4\%     & \multicolumn{1}{l|}{17.5}       & \multicolumn{1}{l|}{15.5}      &    -11.6\%     & \multicolumn{1}{l|}{45.7}       & \multicolumn{1}{l|}{44.0}      &  -3.7\%         \\ \hline
Go                & \multicolumn{1}{l|}{23.4}       & \multicolumn{1}{l|}{22.2}      &    -5.0\%     & \multicolumn{1}{l|}{24.7}       & \multicolumn{1}{l|}{23.4}      &    -5.1\%     & \multicolumn{1}{l|}{49.5}       & \multicolumn{1}{l|}{48.0}      &  -2.9\%         \\ \hline
JS/TS             & \multicolumn{1}{l|}{15.3}       & \multicolumn{1}{l|}{14.0}      &    -8.8\%     & \multicolumn{1}{l|}{19.7}       & \multicolumn{1}{l|}{18.3}      &    -6.9\%     & \multicolumn{1}{l|}{51.9}       & \multicolumn{1}{l|}{50.3}      &  -3.0\%         \\ \hline
Python            & \multicolumn{1}{l|}{18.6}       & \multicolumn{1}{l|}{16.8}      &    -9.4\%     & \multicolumn{1}{l|}{12.8}       & \multicolumn{1}{l|}{10.9}      &    -15.0\%     & \multicolumn{1}{l|}{37.7}       & \multicolumn{1}{l|}{35.9}      &  -4.9\%         \\ \hline
C++               & \multicolumn{1}{l|}{21.0}       & \multicolumn{1}{l|}{20.0}      &    -4.6\%     & \multicolumn{1}{l|}{24.6}       & \multicolumn{1}{l|}{23.6}      &    -4.3\%     & \multicolumn{1}{l|}{51.8}       & \multicolumn{1}{l|}{50.7}      &  -2.1\%         \\ \hline
Average                                 & \multicolumn{1}{l|}{20.0}   & \multicolumn{1}{l|}{18.6}  & \multicolumn{1}{l|}{-7.0\%}  & \multicolumn{1}{l|}{19.9}   & \multicolumn{1}{l|}{18.3}  & \multicolumn{1}{l|}{-8.6\%}  & \multicolumn{1}{l|}{47.3}   & \multicolumn{1}{l|}{45.8}  & \multicolumn{1}{l|}{-3.3\%}  \\ \hline
\end{tabular}
\end{table*}

\section{Discussions and Future Directions}

\begin{figure*}[htb]
	\centering
        \vspace{-1mm}
	\includegraphics[width=1.4\columnwidth]{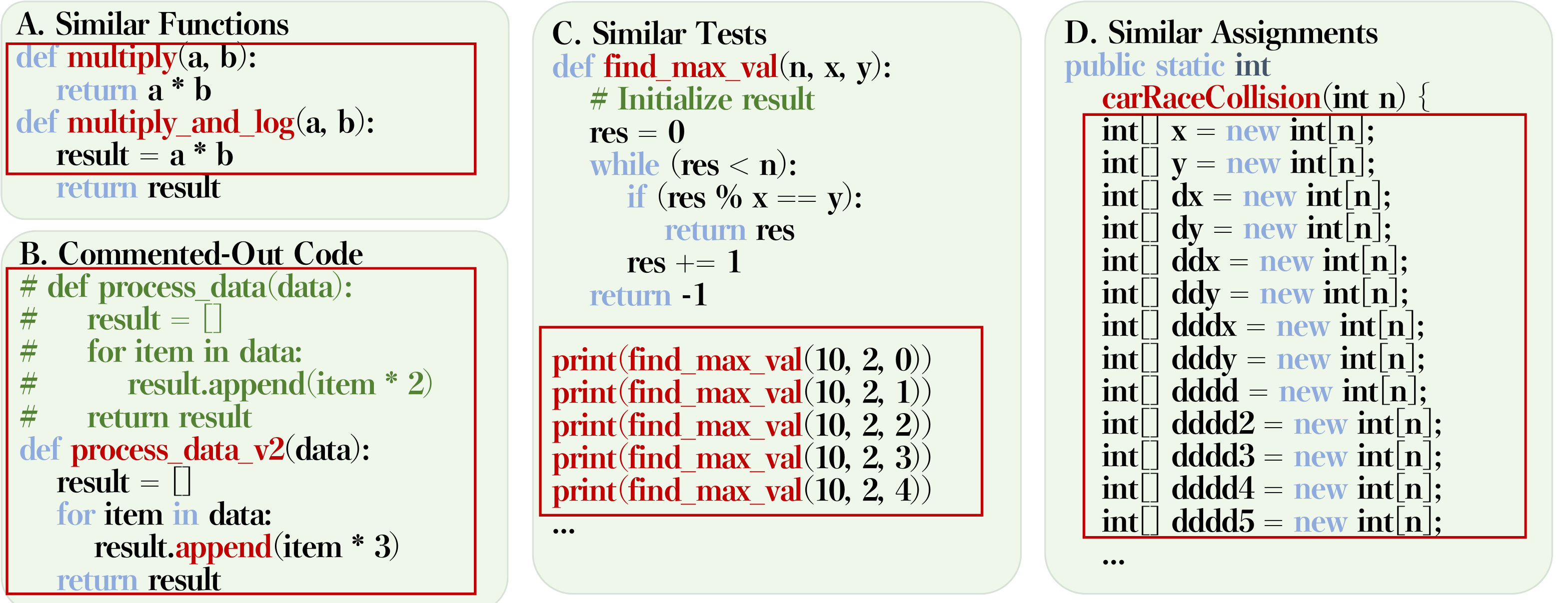}
	\caption{Examples of Repetitive Patterns in Training Corpus of Code LLMs}
	\label{fig:repetition cases}
        \vspace{-1mm}
\end{figure*}

In this section, we explore the distinct characteristics and prevalence of repetition in code generation compared to general text, potential causes, and future directions for mitigation.

\textbf{Repetition is notably more severe in code than in general text due to programming's inherent requirements.} Code frequently involves repetitive patterns, such as repeated variable usage, API calls, and similar conditions or loops. Research~\cite{DBLP:conf/nips/SuLWYKC22} shows that the rep-2, rep-3, and rep-4 values for human-written text are 3.92, 0.88, and 0.28, whereas for code generation tasks, these values are 9.1, 3.3, and 1.7, indicating a much higher prevalence of repetition. 

\textbf{Code cloning, where developers copy and modify existing code, exacerbates repetition in code corpora.} Code LLMs trained on such data learn these repetitive patterns, leading to more significant repetition issues. We found that some of these issues arise from the quality of the pretraining corpus, including prevalent code clones and unrefactored code. Figure~\ref{fig:repetition cases} highlights examples from the training corpus:

\begin{itemize}
    \item Similar Functions: Minor modifications to copied functions lead to repeated patterns in training data.
\item Commented-out Code: Old versions of code left commented out contribute to repetition when the code is later rewritten.
\item Similar Tests: Test cases with slight variations add to repetitive patterns in the training data.
\item Similar Assignments: Repeated variable assignments due to similar logic structures.
\end{itemize}

\textbf{Addressing repetition issues requires high-quality training data.} Future work should focus on assessing how training data quality affects code LLMs and developing methods to address these issues. Retraining with improved data could enhance performance. Additionally, continue training and alignment with human annotations or detection tools could help further reduce code generation repetition.

\section{Threats to Validity}
\parabf{Internal Threats.}
For internal validity, we used public versions of each model per official guidelines to prevent implementation issues. Prompts were standardized across experiments, and greedy decoding was used to reduce randomness. Repetition mitigation techniques were applied using official implementations and tested through controlled experiments. Detailed methods ensure transparency and reproducibility, allowing others to verify and build on our work.

\parabf{External Threats.}
The generalizability of our findings may be limited by the specific models and datasets used in our study, potentially affecting their applicability to other programming environments or real-world applications. To address this, we tested a diverse range of code LLMs and datasets, including different model sizes and training approaches. Additionally, we conducted experiments in real industrial settings (Section \ref{sec:eval:rq5}) to enhance the robustness and applicability of our conclusions. All experimental results are available in our replication package for further verification and validation.

\section{Conclusion}

In this study, we addressed the critical issue of repetition in code generated by LLMs. Through comprehensive quantitative and qualitative analyses, we revealed the significant prevalence and diverse patterns of repetition across various state-of-the-art code LLMs. Based on these findings, we developed DeRep, a rule-based technique designed to detect and mitigate repetition in generated code. Our extensive evaluation, including experiments in real industrial settings, demonstrated that DeRep effectively reduces repetition and enhances overall code quality, meeting real-time performance requirements. This work underscores the importance of addressing repetition in code generation and offers a practical solution to improve the reliability and efficiency of LLM-based code generation tools. Future research will explore further refinement of DeRep and its application to broader programming languages and contexts.

\bibliographystyle{IEEEtran}
\balance
\bibliography{ref}

\end{document}